\PassOptionsToPackage{numbers}{natbib} % citeproc handles citation formatting; prevent natbib author-year conflict
\documentclass[fleqn,10pt]{wlpeerj}

\usepackage{fontspec}

\usepackage{longtable}
\usepackage{array}
\usepackage{etoolbox}

\makeatletter
\patchcmd\longtable{\par}{\if@noskipsec\mbox{}\fi\par}{}{}
\makeatother

\IfFileExists{footnotehyper.sty}{\usepackage{footnotehyper}}{\usepackage{footnote}}
\makesavenoteenv{longtable}

\makeatletter
\newsavebox\pandoc@box
\newcommand*\pandocbounded[1]{% scales image to fit in text height/width
  \sbox\pandoc@box{#1}%
  \Gscale@div\@tempa{\textheight}{\dimexpr\ht\pandoc@box+\dp\pandoc@box\relax}%
  \Gscale@div\@tempb{\linewidth}{\wd\pandoc@box}%
  \ifdim\@tempb\p@<\@tempa\p@\let\@tempa\@tempb\fi%
  \ifdim\@tempa\p@<\p@\scalebox{\@tempa}{\usebox\pandoc@box}%
  \else\usebox{\pandoc@box}%
  \fi%
}
\def\fps@figure{htbp}
\makeatother

\providecommand{\tightlist}{%
  \setlength{\itemsep}{0pt}\setlength{\parskip}{0pt}}

\usepackage{color}
\usepackage{fancyvrb}

\DefineVerbatimEnvironment{Highlighting}{Verbatim}{commandchars=\\\{\}}
\newenvironment{Shaded}{}{}

\newcommand{\NormalTok}[1]{#1}

\NewDocumentCommand\citeproctext{}{}

\makeatletter
 \let\@cite@ofmt\@firstofone
 \def\@biblabel#1{}
 \def\@cite#1#2{{#1\if@tempswa , #2\fi}}
\makeatother
\newlength{\cslhangindent}
\newlength{\csllabelwidth}
\newenvironment{CSLReferences}[2]
 {\begin{list}{}{%
  \setlength{\itemindent}{0pt}
  \setlength{\leftmargin}{0pt}
  \setlength{\parsep}{0pt}
  \ifodd #1
   \setlength{\leftmargin}{\cslhangindent}
   \setlength{\itemindent}{-1\cslhangindent}
  \fi
  \setlength{\itemsep}{#2\baselineskip}}}
 {\end{list}}

\AtBeginEnvironment{longtable}{%
  \scriptsize%
  \setlength{\tabcolsep}{2pt}%
}

\raggedright
\usepackage{hyperref}
\hypersetup{
  pdftitle={Source-Known Identifiers: A Three-Tier Identity System for Distributed Applications},
  pdfauthor={Duran Serkan Kılıç},
  hidelinks,
  pdfcreator={LaTeX via Pandoc}}

\title{Source-Known Identifiers: A Three-Tier Identity System for
Distributed Applications}
\author[1]{Duran Serkan Kılıç}
\affil[1]{Independent Researcher, Ankara, Türkiye}
\corrauthor[1]{Duran Serkan Kılıç}{duran.serkan@outlook.com}

\begin{abstract}
Distributed applications use identifiers across database storage,
trusted communication, and external access. These roles require
different combinations of storage efficiency, chronological sortability,
origin metadata embedding, zero-lookup verification, metadata
confidentiality, and multi-century addressability. The identifier
schemes compared in this paper, including Universally Unique Identifier
(UUID) versions 4 and 7, do not individually provide all six properties.
We present Source-Known Identifiers, a three-tier identity system that
distributes these properties across representations of one entity
identity, connected through deterministic transformations. A 64-bit
Source-Known ID (SKID) provides a timestamp-ordered database key. A
128-bit Source-Known Entity ID (SKEID) adds entity type, epoch, and a
keyed message authentication code (MAC) to the SKID, enabling integrity
and origin metadata checks within a shared-key trust domain without
record lookups. Secure SKEID encrypts this representation to conceal
metadata while retaining linkability of repeated identifiers. Collision
checks prevent generated ciphertext from being accepted as plaintext.
Evaluation combines specification and threat analysis, implementation
tests, and microbenchmarks of the C\#/.NET 10 reference implementation.
Within sequence capacity on Apple M2 hardware, SKEID and Secure SKEID
generation averaged 207.28 ± 3.10 ns and 224.62 ± 3.90 ns, respectively,
compared with 361.94 ± 1.94 ns for UUID Version 4 and 384.11 ± 1.95 ns
for UUID Version 7. ± values indicate 99.9\% confidence interval
margins. Under saturation, generation was slower than the UUID
baselines. These results support the feasibility of deriving SKEIDs from
64-bit Source-Known IDs on demand.
\end{abstract}

\begin{document}

\flushbottom
\maketitle
\thispagestyle{empty}

\raggedbottom

\newpage

\section{Introduction}\label{introduction}

\subsection{Problem Statement}\label{problem-statement}

Distributed applications use identifiers as database keys, internal
references, and externally visible resource handles. These roles impose
different requirements on storage size, ordering, integrity
verification, and metadata exposure. This paper investigates how
deterministic transformations can provide different representations of
one entity identity across these boundaries.

An \emph{entity} is a domain object with a distinct identity that
persists across its lifecycle (Evans, 2003). Its \emph{entity
identifier} distinguishes it from other entities of the same type. While
this terminology is established in Domain-Driven Design, the underlying
requirement of uniquely and persistently identifying records across
distributed systems is universal regardless of architectural style,
design patterns, or definitions. Specifically, modern distributed
applications require entity identifiers that serve multiple roles
simultaneously such as database primary keys, inter-application
correlation tokens, and externally visible resource handles. However,
existing identifier schemes force a choice between conflicting
properties.

Database sequences offer compact storage (4 or 8 bytes) and natural
ordering but expose creation patterns when passed to external consumers.
UUID Version 4 (Davis, Peabody \& Leach, 2024) provides 122 bits of
randomness for probabilistic uniqueness but sacrifices ordering. UUID
Version 7 (Davis, Peabody \& Leach, 2024) restores time-ordering with a
48-bit Unix timestamp but exposes creation patterns. Snowflake-style
identifiers (Twitter Engineering, 2012) embed timestamp and worker
topology in 64 bits but again expose creation patterns and lose
multi-century addressability.

None of these schemes provides built-in verification of identifier
integrity and authenticated origin metadata without a lookup. Exposing
chronological order also reveals information that may need to remain
confidential outside trusted environments.

A multi-tier identifier system can address these challenges by providing
different identifier properties for different trust levels. The proposed
system targets the following six properties across its tiers:

\begin{enumerate}
\def\labelenumi{\arabic{enumi}.}
\tightlist
\item
  Storage efficiency
\item
  Chronological sortability
\item
  Origin metadata embedding
\item
  Zero-lookup verification
\item
  Confidentiality for external consumers
\item
  Multi-century addressability
\end{enumerate}

This study asks under what deployment and cryptographic assumptions
deterministic transformations can distribute the six target properties
across representations of a single entity identity at different trust
boundaries, and how generation costs across the multi-tier system
compare with those of UUID versions 4 and 7. These UUID versions were
selected as standardized baselines defined in RFC 9562, representing
random and time-ordered identifiers, respectively.

\subsection{Motivation}\label{motivation}

At the persistence tier, B-tree index performance favors sequential or
time-ordered primary keys (Davis, Peabody \& Leach, 2024, secs. 2.1,
6.11, and 6.13). However, exposing these identifiers through external
APIs can make enumeration and exploitation of Insecure Direct Object
Reference (IDOR) vulnerabilities easier (MITRE Corporation, 2026a).
Exposure can also enable adversarial inference of generation velocity
and record volume (the German Tank Problem (Ruggles \& Brodie, 1947)).
As a workaround, the dual-identifier pattern with an integer primary key
alongside a random UUID alternate key compromises storage efficiency.

The full cost of such workarounds extends beyond the identifier columns
themselves. For instance, a conventional schema using an auto-increment
primary key (8 bytes), a UUID external identifier (16 bytes), and a
\texttt{created\_at} timestamp (8 bytes) requires 32 bytes of column
data per record. In a fully indexed schema, each column demands a
separate B-tree index, resulting in three indexes and their associated
maintenance operations such as vacuum, reindex, and statistics
collection. An identifier that embeds a timestamp and derives a
UUID-compatible external representation deterministically at application
runtime consolidates all three concerns into a single 8-byte primary key
column with a single index, if the embedded generation timestamp
satisfies the application's \texttt{created\_at} requirement. Under this
fully indexed baseline, this represents a 75\% reduction in raw column
data across these three fields and a reduction from three single-column
indexes to one. Total disk footprint savings depend on storage-engine
tuple overhead (such as row headers and alignment padding) and page fill
factors. Additional composite indexes involving these fields would
amplify the savings further.

Where downstream applications require identifier authenticity
verification (MITRE Corporation, 2026b), an embedded MAC enables
verification of identifier integrity without a query or cache lookup for
the corresponding record. This eliminates record-lookup I/O for that
verification step.

Time-ordered identifiers (Davis, Peabody \& Leach, 2024) can also
support historical analysis when records from independent generators are
combined. Their usefulness depends on timestamp precision, clock
synchronization, and consistent epoch interpretation. When data from
independent generators must be merged years or decades after creation,
the identifier scheme must preserve consistent ordering beyond the
operational lifespan of individual system components. This is especially
critical for any system requiring long-term data retention and
historical analysis.

These requirements motivate a multi-tier identifier system where a
single entity identity is represented differently across trust
boundaries.

\subsection{Prior Work}\label{prior-work}

Source-Known Identifiers build on established techniques for structuring
and protecting identifiers. Snowflake packs timestamps, topology, and
sequence counters into 64-bit identifiers (Twitter Engineering, 2012),
while UUIDv8 permits application-defined fields within a standard
128-bit layout (Davis, Peabody \& Leach, 2024, sec. 5.8). These provide
precedents for SKID's topology fields and SKEID's metadata layout.

The cryptographic foundations are also established. MACs provide
integrity verification, and authenticated encryption combines
confidentiality with authenticity (Song et al., 2006; Harkins, 2008;
McGrew, 2008).

Bellare and Rogaway's encode-then-encipher framework shows how checking
embedded redundancy after deciphering can provide authenticity under a
strong pseudorandom-permutation assumption (Bellare \& Rogaway, 2000,
Theorem 2). SKEID incorporates an embedded MAC, and Secure SKEID follows
this principle by encrypting the structured representation and verifying
it after decryption.

For external use, Khuong (2022) describes using AES to convert a
structured 128-bit key into an opaque identifier of the same size. The
article also discusses cycle walking and a specialized Feistel
construction for producing encrypted identifiers with fixed format bits,
including those required by UUIDv8. As discussed in Future Work,
preserving UUID version and variant bits may be unnecessary where the
complete application path accepts opaque 128-bit identifiers.

IDMask combines reversible identifier encryption with forgery checks
(Favre, 2023). Its default deterministic 64-bit mode encrypts the
identifier and 64 bits of validation redundancy in one AES block, then
adds a version byte to produce a 17-byte output. Its default
deterministic 128-bit mode adds an 8-byte MAC and a version byte to the
encrypted identifier, producing 25 bytes. The MAC in the 128-bit mode
remains visible outside the ciphertext. In both modes, the version byte
is XOR-obfuscated with the first ciphertext byte and can be recovered
without the key. Both outputs exceed 128 bits. IDMask is excluded from
the comparison tables because it protects application-supplied
identifiers, whereas the tables focus on identifier-generation schemes.

Source-Known Identifiers bring these techniques together across three
representations of one identity, combining compact storage,
authenticated entity-type and origin metadata, and encryption for
external use. Conversion from SKEID to Secure SKEID preserves the
128-bit size. Collision-guard and backward-verification mechanisms
distinguish the plaintext and encrypted representations.

\newpage

\subsection{Contributions}\label{contributions}

The contributions concern the three-tier architecture and its generation
and parsing mechanisms. AES-256 and BLAKE3 are established primitives.
The design combines them with embedded metadata, a collision guard, and
backward verification while preserving the 128-bit size when converting
SKEID to Secure SKEID.

\begin{enumerate}
\def\labelenumi{\arabic{enumi}.}
\tightlist
\item
  \textbf{Design of a three-tier identity system} that satisfies all six
  desired identifier properties across its three tiers, aligned with
  trust boundaries (database, trusted internal, and external).
\item
  \textbf{A defense-in-depth security architecture} combining AES-256
  encryption, BLAKE3 keyed MAC, marker byte detection, epoch validation,
  and optional entity type and application ID matching.
\item
  \textbf{A collision guard for plaintext-to-encrypted identifier
  conversion} that checks candidate ciphertext against plaintext
  acceptance criteria and, on collision, recomputes the MAC and
  re-encrypts successive variants. Successfully generated ciphertext
  cannot be accepted as plaintext.
\item
  \textbf{Cryptographic backward verification} that enables the parser
  to reconstruct and verify every preceding collision under the
  authenticated key pair before accepting a non-default variant. This
  mechanism verifies that every preceding variant would have triggered
  the collision guard, without trusting that an input was produced by
  the generator.
\item
  \textbf{Zero-expansion plaintext-to-encrypted identifier conversion}
  that combines embedded metadata, a truncated MAC, and collision-guard
  variants within the same 128-bit representation using AES's
  length-preserving block permutation. By comparison, AES-SIV and the
  AES-GCM profiles in RFC 5116 each add 16 bytes to the plaintext
  length, excluding any separately transmitted nonce (Harkins, 2008,
  sec. 2.6; McGrew, 2008, sec. 5.1 and 5.2).
\item
  \textbf{An epoch addressing system} embedded in the plaintext SKEID's
  UUIDv8-compatible layout that spans approximately 17,421 years of
  total coverage through 256 configurable epochs of \(2^{31}\) seconds
  each (approximately 68 years per epoch).
\item
  \textbf{A reference implementation} in C\#/.NET 10 with open-source
  code, integration and unit test coverage, and BenchmarkDotNet
  performance data demonstrating lower mean generation times than
  standard UUID generation within sequence capacity, with separate
  saturation measurements of backpressure.
\end{enumerate}

\subsection{Related Work}\label{related-work}

Distributed identifier schemes have evolved considerably since the
original UUID specification. RFC 9562 (Davis, Peabody \& Leach, 2024)
standardized UUID versions 1, 3--8, with Version 7 introducing
timestamp-ordered UUIDs that address the B-tree fragmentation problems
of random Version 4 UUIDs. UUID Version 7 uses a 48-bit Unix timestamp
prefix followed by random bits, providing millisecond-precision ordering
and global uniqueness.

The comparison includes the alternatives examined during the development
of RFC 9562 including Snowflake, ULID, KSUID, and the original CUID
(Davis, Peabody \& Leach, 2024, sec. 2.1). These schemes illustrate
different choices in identifier size, ordering, timestamp precision, and
generator metadata. We include CUID2 as CUID's successor, UUID V4 and V7
as standardized random and time-ordered baselines, and database
sequences as the compact storage baseline. This selection covers
relevant design trade-offs without reproducing the RFC's full survey.

Twitter's Snowflake architecture (Twitter Engineering, 2010, 2012)
pioneered timestamp-prefixed, worker-partitioned 64-bit identifiers for
high-throughput systems. Snowflake identifiers embed a 41-bit timestamp
with millisecond precision, a 10-bit machine identifier (5-bit
datacenter ID and 5-bit worker ID), and a 12-bit sequence number,
enabling approximately 4,096 identifiers per millisecond per machine.
Its custom epoch starts from November 4, 2010 and extends to July 10,
2080 (\textasciitilde69.68 years).

ULIDs (Feerasta, 2019) provide a 128-bit value composed of a 48-bit
timestamp with millisecond precision. Its Unix epoch starts from January
1, 1970 and extends to \textasciitilde10889 CE (\textasciitilde8,919
years). ULIDs are encoded as 26-character Crockford Base32 strings that
sort lexicographically by creation time, targeting environments where
string-based identifiers are standard.

The original CUID (Elliott, 2023) was a collision-resistant identifier
specification that used a k-sortable, timestamp-prefixed structure. It
was deprecated by its author due to security concerns. The CUID
deprecation notice warns that ``all monotonically increasing
(auto-increment, k-sortable), and timestamp-based ids share the security
issues with Cuid'' and further states that ``UUID V6-V8 are also
insecure because they leak information which could be used to exploit
systems or violate user privacy'' (Elliott, 2023). We note that this
warning is valuable but too broad for UUIDv8. Whether UUIDv8 shares
these security issues depends on its application-defined payload (Davis,
Peabody \& Leach, 2024, sec. 5.8).

CUID2, the successor to CUID, takes a fundamentally different approach.
The CUID2 authors observe that ``insecure ids can cause problems in
unexpected ways, including unauthorized user account access,
unauthorized access to user data, and accidental leaks of user's
personal data which can lead to catastrophic effects, even in
innocent-sounding applications like fitness run trackers'' (Elliott,
2026). CUID2 intentionally removed timestamps from the identifier for
security reasons and instead recommends a separate indexed
\texttt{createdAt} column for time-based sorting. CUID2 generates
identifiers by using independent entropy sources then hashing the
concatenation with SHA3. This produces identifiers with strong collision
resistance but relies on probabilistic uniqueness rather than
deterministic construction.

KSUIDs (Segment, 2023) (K-Sortable Unique Identifier) provide a 160-bit
(20-byte) binary value composed of a 32-bit timestamp. It has 1-second
precision, a custom epoch from May 13, 2014 to June 19, 2150, and a
128-bit cryptographically random payload. KSUIDs also have a
27-character Base62 representation. Both binary and text representations
sort lexicographically by their embedded second-resolution timestamps.
Identifiers generated within the same second are ordered by their random
payloads. The 128-bit random payload provides stronger collision
resistance than UUID V4's 122 random bits.

None of these approaches offers integrity verification. Table 1 presents
a detailed feature comparison.

\noindent\begin{minipage}{\linewidth}

\textbf{Table 1:} Feature comparison of identifier schemes.

{\def\LTcaptype{none} % do not increment counter
\begin{longtable}[]{@{}
  >{\raggedright\arraybackslash}p{(\linewidth - 16\tabcolsep) * \real{0.2061}}
  >{\raggedright\arraybackslash}p{(\linewidth - 16\tabcolsep) * \real{0.1679}}
  >{\raggedright\arraybackslash}p{(\linewidth - 16\tabcolsep) * \real{0.0916}}
  >{\raggedright\arraybackslash}p{(\linewidth - 16\tabcolsep) * \real{0.0687}}
  >{\raggedright\arraybackslash}p{(\linewidth - 16\tabcolsep) * \real{0.0840}}
  >{\raggedright\arraybackslash}p{(\linewidth - 16\tabcolsep) * \real{0.0763}}
  >{\raggedright\arraybackslash}p{(\linewidth - 16\tabcolsep) * \real{0.0763}}
  >{\raggedright\arraybackslash}p{(\linewidth - 16\tabcolsep) * \real{0.1298}}
  >{\raggedright\arraybackslash}p{(\linewidth - 16\tabcolsep) * \real{0.0992}}@{}}
\toprule\noalign{}
\begin{minipage}[b]{\linewidth}\raggedright
Feature
\end{minipage} & \begin{minipage}[b]{\linewidth}\raggedright
Source-Known IDs
\end{minipage} & \begin{minipage}[b]{\linewidth}\raggedright
UUID V4
\end{minipage} & \begin{minipage}[b]{\linewidth}\raggedright
UUID V7
\end{minipage} & \begin{minipage}[b]{\linewidth}\raggedright
Snowflake
\end{minipage} & \begin{minipage}[b]{\linewidth}\raggedright
ULID
\end{minipage} & \begin{minipage}[b]{\linewidth}\raggedright
CUID2
\end{minipage} & \begin{minipage}[b]{\linewidth}\raggedright
KSUID
\end{minipage} & \begin{minipage}[b]{\linewidth}\raggedright
DB Sequence
\end{minipage} \\
\midrule\noalign{}
\endhead
\bottomrule\noalign{}
\endlastfoot
Chronological ordering & 250ms & No & ms & ms & ms & No & second &
Numeric \\
Embedded timestamp & Yes & No & Yes & Yes & Yes & No & Yes & No \\
Origin metadata embedding & Yes & No & No & Partial & No & No & No &
No \\
Multi-century addressing & Yes (SKEID) & N/A & Yes & No & Yes & N/A &
Partial & N/A \\
Entity type & Yes (8-bit) & No & No & No & No & No & No & No \\
Integrity check & Yes (BLAKE3 MAC) & No & No & No & No & No & No & No \\
Confidentiality & Encryption (AES-256) & Randomness & No & No & No &
Hashing & No & No \\
Zero-lookup verification & Yes & No & No & No & No & No & No & No \\
DB primary key size & 8 B (SKID) & 16 B & 16 B & 8 B & 16 B & 24 chars &
20 B & 4/8 B \\
External ID size & 16 B (Secure SKEID) & 16 B & 16 B & 8 B & 26 chars &
24 chars & 27 chars & 4/8 B \\
UUID compatible & Yes (SKEID, UUID V8) & Yes & Yes & No & No & No & No &
No \\
\end{longtable}
}

\end{minipage}

\section{Methods}\label{methods}

Portions of this text were previously published as part of an arXiv
preprint (Kılıç, 2026a).

The proposed identifier system is evaluated through specification
analysis, implementation tests, performance measurement, and threat
analysis. The evaluation examines representation size and ordering,
identity preservation and persistence behavior, execution time and
managed allocation, sequence-capacity behavior, and security properties
under the stated deployment and cryptographic assumptions.

Performance evaluation uses microbenchmarks of generation, parsing, and
conversion, compares generation costs with UUID versions 4 and 7, and
examines backpressure through separate saturation measurements.
Database-engine performance is not evaluated, and the security analysis
does not constitute a formal compositional proof.

\subsection{Three-Tier Identity Model}\label{three-tier-identity-model}

Source-Known Identifiers represent entity identity through a three-tier
identity model that spans database storage, trusted internal
communication (in-process or inter-process), and external access.

\begin{enumerate}
\def\labelenumi{\arabic{enumi}.}
\item
  \textbf{Database Tier:} A 64-bit Source-Known ID (SKID) serves as the
  primary key. It embeds a 250-millisecond-precision timestamp,
  application topology (application identifier, instance identifier),
  and a per-entity-type sequence counter, providing natural ordering and
  compact B-tree indexing. Entity-type and epoch context are supplied by
  the surrounding schema and configuration, for example through the
  table name and the dataset's configured epoch.
\item
  \textbf{Trusted Environment Tier:} A 128-bit Source-Known Entity ID
  (SKEID) encodes the SKID, the entity type, epoch metadata, a keyed MAC
  for integrity, and RFC 9562 Version 8 compatible markers. Applications
  can verify identifier integrity and check the asserted origin and
  entity type without a database lookup.
\item
  \textbf{External Tier:} A 128-bit Secure Source-Known Entity ID
  (Secure SKEID) encrypts the entire SKEID block using AES-256,
  concealing embedded metadata from untrusted consumers while retaining
  the conventional UUID text format (\texttt{8-4-4-4-12} hexadecimal
  digits; see Limitation 6).
\end{enumerate}

The three tiers are not three separate identifiers but one entity
identity projected for different trust boundaries. The database stores
the compact SKID, trusted applications derive the SKEID on demand, and
external consumers receive the Secure SKEID. Deterministic bidirectional
transformations between tiers allow any representation to be converted
to any other, given the required entity-type, epoch context and the
appropriate cryptographic keys. These transformations require no I/O or
external lookup. Figure 1 illustrates the data transformations between
the three tiers.

\begin{verbatim}
   Database              Trusted Internal          External / Public
  ┌──────────┐          ┌───────────────┐          ┌────────────────┐
  │ SKID     │ Generate │ SKEID         │ ToSecure │ Secure SKEID   │
  │ Sortable │─────────▶│ BLAKE3 MAC    │─────────▶│ AES Encrypted  │
  │ 64-bit   │◀─────────│ 128-bit       │◀─────────│ 128-bit        │
  └──────────┘  Parse   └───────────────┘ ToPlain  └────────────────┘
\end{verbatim}

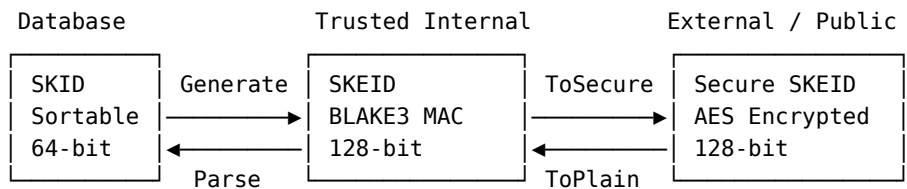
\textbf{Figure 1:} Three-tier identity model: bidirectional data
transformations.

\subsection{Source-Known ID (SKID) 64-bit
Specification}\label{source-known-id-skid-64-bit-specification}

\subsubsection{Bit Layout}\label{bit-layout}

A SKID is a 64-bit signed integer with the following field layout
(Figure 2), packed most-significant bit first.

\begin{verbatim}
 0                   1                   2                   3
 0 1 2 3 4 5 6 7 8 9 0 1 2 3 4 5 6 7 8 9 0 1 2 3 4 5 6 7 8 9 0 1
+-+-+-+-+-+-+-+-+-+-+-+-+-+-+-+-+-+-+-+-+-+-+-+-+-+-+-+-+-+-+-+-+
|S|                    Timestamp (T - 32 bits)                  |
+-+-+-+-+-+-+-+-+-+-+-+-+-+-+-+-+-+-+-+-+-+-+-+-+-+-+-+-+-+-+-+-+
|T| App ID (7)  | Inst (6)  |         Sequence (18 bits)        |
+-+-+-+-+-+-+-+-+-+-+-+-+-+-+-+-+-+-+-+-+-+-+-+-+-+-+-+-+-+-+-+-+
\end{verbatim}

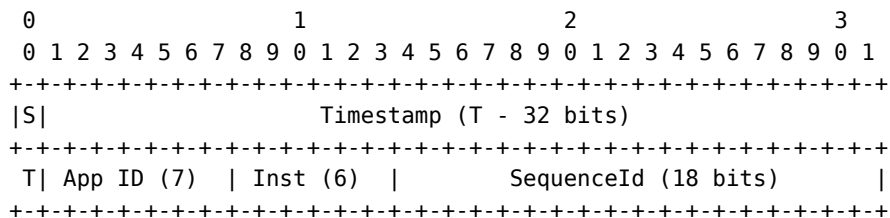
\textbf{Figure 2:} SKID 64-bit field layout (IETF RFC diagram
convention).

Total: 1 (sign) + 32 (timestamp) + 7 (app ID) + 6 (app instance ID) + 18
(sequence counter) = 64 bits.

Table 2 details the corresponding fields.

\textbf{Table 2:} SKID field definitions.

{\def\LTcaptype{none} % do not increment counter
\begin{longtable}[]{@{}
  >{\raggedright\arraybackslash}p{(\linewidth - 6\tabcolsep) * \real{0.1597}}
  >{\raggedright\arraybackslash}p{(\linewidth - 6\tabcolsep) * \real{0.0672}}
  >{\raggedright\arraybackslash}p{(\linewidth - 6\tabcolsep) * \real{0.0756}}
  >{\raggedright\arraybackslash}p{(\linewidth - 6\tabcolsep) * \real{0.6975}}@{}}
\toprule\noalign{}
\begin{minipage}[b]{\linewidth}\raggedright
Field
\end{minipage} & \begin{minipage}[b]{\linewidth}\raggedright
Bit(s)
\end{minipage} & \begin{minipage}[b]{\linewidth}\raggedright
Width
\end{minipage} & \begin{minipage}[b]{\linewidth}\raggedright
Purpose
\end{minipage} \\
\midrule\noalign{}
\endhead
\bottomrule\noalign{}
\endlastfoot
Sign / Epoch half & 63 & 1 bit & Epoch half indicator for smooth epoch
transitions \\
Timestamp & 62--31 & 32 bits & 250ms ticks relative to the epoch half
(approximately 34 years per half) \\
App ID & 30--24 & 7 bits & Application identifier (maximum 127) \\
App Instance ID & 23--18 & 6 bits & Instance discriminator (maximum
63) \\
Sequence counter & 17--0 & 18 bits & Per application instance and
entity-type monotonic counter \\
\end{longtable}
}

The sign bit controls epoch half selection. When set to 1, the resulting
value is negative, covering the first approximately 34 years of the
epoch. When set to 0, the value becomes nonnegative, extending coverage
for the second approximately 34 years. Together, the two halves span
approximately 68 years per epoch while preserving monotonic sort order
in signed 64-bit representation, since negative values sort before
nonnegative values. This timestamp-leading layout produces a
chronologically sortable SKID optimized for internal database indexing
and ordering.

The 32-bit timestamp field stores unsigned 250-millisecond ticks elapsed
since the start of the current epoch half. Sub-second ordering at 250ms
granularity eliminates coarse-grained temporal ambiguity while
preserving throughput within a compact 64-bit layout.

\subsubsection{Sequence Capacity}\label{sequence-capacity}

For each application/entity-type pair on one instance, the 18-bit
counter provides a \textbf{sequence capacity} of \(2^{18} = 262{,}144\)
identifiers per tick. At four ticks per second, this permits
\(4 \times 262{,}144 = 1{,}048{,}576\) identifiers per second. Across an
application's 64 instances, the \textbf{aggregate sequence capacity} for
that pair is \(64 \times 1{,}048{,}576 = 67{,}108{,}864\) identifiers
per second.

Implementations must reset the sequence when the timestamp advances. If
the sequence capacity is exhausted within a tick, the generator must
wait for the next tick before issuing new identifiers for that
application/entity-type pair on that instance, applying backpressure to
maintain uniqueness guarantees.

\subsubsection{Generation Algorithm}\label{generation-algorithm}

The SKID generation procedure operates as follows. Given
\texttt{entityType}, \texttt{appId}, \texttt{appInstanceId}, and a
configured \texttt{epoch}:

\begin{enumerate}
\def\labelenumi{\arabic{enumi}.}
\tightlist
\item
  Compute \texttt{elapsedTicks} as the floor of 250-millisecond ticks
  elapsed since the epoch (four ticks per second).
\item
  Assert that the elapsed ticks are within the epoch range (0 to
  \(2^{33} - 1\) ticks, covering both epoch halves).
\item
  Determine the epoch half:

  \begin{enumerate}
  \def\labelenumii{\arabic{enumii}.}
  \tightlist
  \item
    If \texttt{elapsedTicks} \(< 2^{32}\), the current half is the first
    half (sign bit = 1, producing a negative SKID). Otherwise, the
    current half is the second half (sign bit = 0, producing a
    nonnegative SKID).
  \item
    Compute \texttt{timestamp} = \texttt{elapsedTicks} mod \(2^{32}\)
    (equivalently, \texttt{elapsedTicks} AND \(2^{32} - 1\)).
  \end{enumerate}
\item
  Obtain the next sequence value from the per-entity-type atomic counter
  (resetting on timestamp change).
\item
  Pack fields into a 64-bit signed integer: sign bit at position 63,
  timestamp at positions 62--31, \texttt{appId} at positions 30--24,
  \texttt{appInstanceId} at positions 23--18, and sequence counter at
  positions 17--0.
\item
  Return the packed value.
\end{enumerate}

\subsubsection{Clock Drift Protection}\label{clock-drift-protection}

The system handles backward time jumps at two levels. Minor drifts
freeze the timestamp until the wall clock catches up, allowing the
sequence counter to continue advancing within the frozen tick. The
reference implementation uses a 5-second freeze threshold. Drifts at or
beyond this threshold are considered critical. Subsequent generation
requests fail, and application shutdown is requested. Restarting the
application and assigning a new app instance ID are responsibilities of
an external configuration or coordination service (see Limitation 2).
This mechanism maintains nondecreasing generation timestamps within a
running process. A new app instance ID is used to prevent collisions
across application restarts and does not preserve chronological
monotonicity across a restart after clock rollback.

\newpage

\subsection{Source-Known Entity ID (SKEID) 128-bit
Specification}\label{source-known-entity-id-skeid-128-bit-specification}

\subsubsection{Byte Layout}\label{byte-layout}

An SKEID occupies 16 bytes (128 bits), structured as shown in Figure 3.

\begin{verbatim}
Byte:  0  1  2  3  4  5  6  7  8  9 10 11 12 13 14 15
     +--+--+--+--+--+--+--+--+--+--+--+--+--+--+--+--+
     |EP| SKID Upper|S0|VE|ET|VA|S1|S2|S3|    MAC    |
     +--+--+--+--+--+--+--+--+--+--+--+--+--+--+--+--+
\end{verbatim}

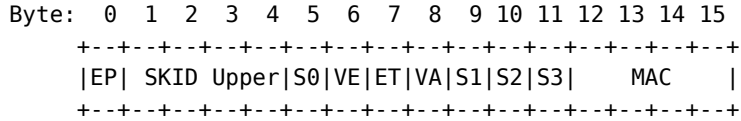
\textbf{Figure 3:} SKEID 16-byte field layout (big-endian, RFC 9562
network byte order).

Table 3 details the byte-level layout of the SKEID fields.

\textbf{Table 3:} SKEID byte layout.

{\def\LTcaptype{none} % do not increment counter
\begin{longtable}[]{@{}
  >{\raggedright\arraybackslash}p{(\linewidth - 6\tabcolsep) * \real{0.1889}}
  >{\raggedright\arraybackslash}p{(\linewidth - 6\tabcolsep) * \real{0.1000}}
  >{\raggedright\arraybackslash}p{(\linewidth - 6\tabcolsep) * \real{0.1000}}
  >{\raggedright\arraybackslash}p{(\linewidth - 6\tabcolsep) * \real{0.6111}}@{}}
\toprule\noalign{}
\begin{minipage}[b]{\linewidth}\raggedright
Field
\end{minipage} & \begin{minipage}[b]{\linewidth}\raggedright
Byte(s)
\end{minipage} & \begin{minipage}[b]{\linewidth}\raggedright
Width
\end{minipage} & \begin{minipage}[b]{\linewidth}\raggedright
Purpose
\end{minipage} \\
\midrule\noalign{}
\endhead
\bottomrule\noalign{}
\endlastfoot
Epoch & 0 & 8 bits & Epoch index (highest lexicographic priority) \\
SKID upper half & 1--4 & 32 bits & SKID bits 63--32, sign-toggled for
lexicographic order \\
SKID low byte 0 & 5 & 8 bits & MSB of SKID lower half (timestamp LSB,
app ID) \\
Version marker & 6 & 8 bits & \texttt{0x8D} (UUID V8, RFC 9562 Section
5.8 octet 6) \\
Entity Type & 7 & 8 bits & Domain entity classification \\
Variant marker & 8 & 8 bits & \texttt{0x8D} (RFC 9562 Section 4.1 octet
8 variant) \\
SKID low bytes & 9--11 & 24 bits & Remaining SKID lower half \\
MAC & 12--15 & 32 bits & BLAKE3 keyed MAC \\
\end{longtable}
}

All fields are encoded in big-endian (network byte order) per RFC 9562.
The epoch byte occupies byte 0 to ensure that higher epoch values sort
lexicographically after lower epoch values, regardless of timestamp. The
upper SKID half at bytes 1--4 is sign-toggled (XOR with
\texttt{0x80000000}) before encoding. This converts the signed SKID sort
order (negative before nonnegative) to unsigned byte sort order,
preserving chronological ordering in lexicographic comparisons across
both epoch halves. The most significant byte of the lower SKID half
occupies byte 5, immediately after the upper half, ensuring that the
timestamp least-significant bit and leading application topology bits
participate in lexicographic comparison before the entity type at byte
7. The remaining lower SKID half is split around the variant marker at
byte 8, with bytes 9--11 holding the last three bytes. Since the version
marker, entity type, and variant marker are constant for all plain
SKEIDs of a given type, lexicographic comparison of the split lower half
operates correctly.

The marker bytes at positions 6 and 8 serve dual purposes.

\begin{enumerate}
\def\labelenumi{\arabic{enumi}.}
\tightlist
\item
  \textbf{UUID V8 compatibility} makes the SKEID a valid RFC 9562 UUID.
\item
  \textbf{Detection signal} enables the parser to distinguish plaintext
  SKEIDs from encrypted Secure SKEIDs.
\end{enumerate}

\subsubsection{MAC Computation}\label{mac-computation}

Integrity verification relies on BLAKE3 keyed MAC (O'Connor et al.,
2021). The computation procedure clears the four MAC bytes (positions
12--15) to zero, then computes a 4-byte BLAKE3 keyed MAC over the full
16-byte buffer. The resulting 4 bytes are written into the contiguous
MAC positions (bytes 12--15).

BLAKE3 was selected for performance and security margin. In the
BenchmarkDotNet measurements, BLAKE3 keyed hashing with a reused hasher
averaged 81.03 ns on 16-byte inputs, compared with 306.75 ns for
HMAC-SHA-256, approximately 3.79 times faster (see supplementary file
\texttt{supplementary-hash-benchmark-report.html}).

Generated MAC is truncated to 32 bits, offering a \(2^{-32}\) success
probability per guessed tag. This truncation is acceptable within the
defense-in-depth architecture where the MAC is one of multiple
verification layers, not the sole security mechanism (see MAC Truncation
Analysis).

The epoch byte (byte 0) is not cleared before MAC computation. It
participates in the MAC input, adding tamper-resistance for the epoch
field.

\subsubsection{Generation Algorithm}\label{generation-algorithm-1}

Given a 64-bit SKID, epoch, entity type, and MAC key, the SKEID
generation procedure:

\begin{enumerate}
\def\labelenumi{\arabic{enumi}.}
\tightlist
\item
  Allocate a 16-byte buffer initialized to zero.
\item
  Split the SKID into upper and lower 32-bit halves. Toggle the most
  significant bit of the upper half (XOR with \texttt{0x80000000}) and
  write it in big-endian to bytes 1--4.
\item
  Write the most significant byte of the SKID lower half to byte 5.
\item
  Write the epoch byte to byte 0, the version marker \texttt{0x8D} to
  byte 6 (RFC 9562 Section 5.8 octet 6), entity type to byte 7, and
  variant marker to byte 8 (RFC 9562 Section 4.1 octet 8).
\item
  Write the remaining three bytes of the SKID lower half to bytes 9--11.
\item
  Compute the BLAKE3 keyed MAC and write it to bytes 12--15.
\item
  Construct the UUID from the 16-byte buffer, interpreting all fields in
  big-endian (network byte order) per RFC 9562, and return the result.
\end{enumerate}

\subsubsection{Parsing Algorithm}\label{parsing-algorithm}

Given a 16-byte UUID \texttt{entityId} and MAC key \(K_{\mathrm{mac}}\):

\begin{enumerate}
\def\labelenumi{\arabic{enumi}.}
\tightlist
\item
  Convert the UUID to a 16-byte big-endian byte array (network byte
  order, per RFC 9562).
\item
  Check that the epoch byte at position 0 represents a supported epoch
  (\texttt{guidBytes{[}0{]}\ \textless{}=\ MaxSupportedEpoch}) and check
  for exact primary marker bytes (\texttt{0x8D} at position 6,
  \texttt{0x8D} at position 8) in the byte array. If the epoch or
  markers do not match, return INVALID.
\item
  Extract the 32-bit MAC from bytes 12--15 and clear bytes 12--15 to
  zero.
\item
  Compute the 4-byte BLAKE3 keyed MAC over the 16-byte buffer using
  \(K_{\mathrm{mac}}\). If it does not match the extracted MAC, return
  INVALID.
\item
  Extract the epoch byte (byte 0) and entity type (byte 7). Reconstruct
  the 64-bit SKID from the upper half (bytes 1--4, reversing the
  sign-bit toggle with \texttt{0x80000000}) and lower half (byte 5 and
  bytes 9--11).
\item
  Return the extracted SKID, epoch, and entity type as a valid plaintext
  SKEID.
\end{enumerate}

\subsection{Secure SKEID 128-bit
Specification}\label{secure-skeid-128-bit-specification}

\subsubsection{AES-256-ECB Single-Block
Encryption}\label{aes-256-ecb-single-block-encryption}

Table 4 shows the Secure SKEID byte layout after AES-256 encryption.

\textbf{Table 4:} Secure SKEID byte layout.

{\def\LTcaptype{none} % do not increment counter
\begin{longtable}[]{@{}
  >{\raggedright\arraybackslash}p{(\linewidth - 6\tabcolsep) * \real{0.1889}}
  >{\raggedright\arraybackslash}p{(\linewidth - 6\tabcolsep) * \real{0.1000}}
  >{\raggedright\arraybackslash}p{(\linewidth - 6\tabcolsep) * \real{0.1111}}
  >{\raggedright\arraybackslash}p{(\linewidth - 6\tabcolsep) * \real{0.6000}}@{}}
\toprule\noalign{}
\begin{minipage}[b]{\linewidth}\raggedright
Field
\end{minipage} & \begin{minipage}[b]{\linewidth}\raggedright
Byte(s)
\end{minipage} & \begin{minipage}[b]{\linewidth}\raggedright
Width
\end{minipage} & \begin{minipage}[b]{\linewidth}\raggedright
Purpose
\end{minipage} \\
\midrule\noalign{}
\endhead
\bottomrule\noalign{}
\endlastfoot
Ciphertext & 0--15 & 128 bits & AES-256 encrypted SKEID (pseudorandom
bytes) \\
\end{longtable}
}

As a single AES block, the entire SKEID is encrypted. No internal
structure is visible to external consumers. Identical SKEID plaintexts
encrypted under the same key produce identical ciphertexts, allowing
repeated identifiers to be recognized.

This conversion has zero expansion. A 16-byte SKEID becomes a 16-byte
Secure SKEID. By comparison, AES-SIV supports deterministic
authenticated encryption but adds 16 bytes of overhead, producing 32
bytes for a 16-byte plaintext (Harkins, 2008, sec. 2.6 and 4). The
AES-GCM profiles in RFC 5116 likewise add a 16-byte authentication tag,
excluding any separately transmitted nonce (McGrew, 2008, sec. 5.1 and
5.2). Although NIST SP 800-38D permits several tag lengths, all require
additional space, so the complete authenticated output still expands
(Dworkin, 2007, sec. 5.2.1.2). Neither construction can therefore
reversibly carry a SKEID in a self-contained 128-bit output. Zero
expansion does not imply security equivalent to AES-SIV or AES-GCM, or
preservation of UUID version and variant bits (see Limitations 4 and 6).

A Secure SKEID is produced by encrypting the entire 16-byte SKEID
plaintext using AES-256 (National Institute of Standards and Technology,
2023) as a single-block cipher. AES has undergone extensive public
cryptanalysis, reviewed in NIST IR 8319 (Mouha, 2021). Under the
pseudorandom permutation (PRP) assumption with a uniformly chosen key,
AES-256 applied to a single 128-bit block is computationally
indistinguishable from a uniformly random permutation. A pseudorandom
function (PRF) is similarly computationally indistinguishable from a
uniformly random function. The birthday threshold for distinguishing a
random permutation from a random function is approximately \(2^{n/2}\)
queries (Bellare, Krovetz \& Rogaway, 1998, sec. 2.2). For AES with
\(n = 128\), candidate ciphertexts for distinct plaintexts can be
modeled approximately as independent, uniformly random blocks as long as
the query count under a single key remains well below \(2^{64}\).

Because SKEID encryption always operates on exactly one block, the
multi-block weakness of ECB mode does not apply. For a single block, ECB
is mathematically identical to Cipher Block Chaining (CBC) with a zero
initialization vector, as shown in Equation
\eqref{eq:aes-cbc-equivalence}:

\begin{equation}
\label{eq:aes-cbc-equivalence}
C = \operatorname{AES}_{K_{\mathrm{aes}}}(P \oplus \mathbf{0})
  = \operatorname{AES}_{K_{\mathrm{aes}}}(P),
\end{equation}

where \(P\) is the plaintext block and \(\mathbf{0}\) is the all-zero
128-bit block. No nonce is required, and there is no nonce-reuse
vulnerability.

Quantum attacks on AES via Grover's algorithm reduce the effective
security of AES-256 to approximately 128 bits, which still remains
computationally infeasible (Bonnetain, Naya-Plasencia \& Schrottenloher,
2019).

\subsubsection{Key Separation}\label{key-separation}

The three-tier identity system uses separate MAC and encryption keys for
each key-ring entry. A MAC key is used for BLAKE3 keyed MAC (integrity
verification) and an AES key is used for AES-256-ECB (confidentiality).
The two keys must not be the same value. Implementations should derive
both keys from a single master secret using a key derivation function
with domain separation, such as BLAKE3's \texttt{derive\_key} mode with
distinct, hardcoded, globally unique context strings identifying the
application and key purpose (O'Connor et al., 2021, sec. 6.2).

\subsubsection{Generation Algorithm}\label{generation-algorithm-2}

Given a 64-bit SKID, epoch, entity type, and key pair
\((K_{\mathrm{mac}}, K_{\mathrm{aes}})\):

\begin{enumerate}
\def\labelenumi{\arabic{enumi}.}
\tightlist
\item
  Construct the 16-byte plaintext SKEID buffer using the SKEID
  generation algorithm with default variant marker \texttt{0x8D}, and
  compute its MAC using \(K_{\mathrm{mac}}\).
\item
  Encrypt the 16-byte buffer using AES-256-ECB with \(K_{\mathrm{aes}}\)
  to produce candidate ciphertext.
\item
  Check whether the ciphertext triggers the collision guard (Collision
  Guard Mechanism below): if the ciphertext exhibits a supported epoch,
  exact markers \texttt{0x8D} at byte 6 and \texttt{0x8D} at byte 8, and
  a valid MAC under any enabled verification key, regenerate with
  successive variant bytes from \texttt{0x8E} through \texttt{0xBF}
  (recomputing the MAC and re-encrypting with the same key pair) until a
  non-colliding ciphertext is obtained.
\item
  Construct the UUID from the final 16-byte ciphertext in big-endian
  order and return the Secure SKEID.
\end{enumerate}

\subsubsection{Collision Guard
Mechanism}\label{collision-guard-mechanism}

Without the collision guard, there exists a combined approximately
\(2^{-48}\) probability that ciphertext coincidentally matches both the
plaintext marker bytes and produces a valid MAC when interpreted as a
plaintext SKEID. This would cause a Secure SKEID to be misclassified as
a plaintext SKEID with incorrect data.

The collision guard eliminates this edge case for the verification keys
enabled during generation. When the ciphertext passes plaintext marker,
epoch, and MAC checks under any enabled key, the plaintext is
regenerated with successive variant bytes from \texttt{0x8E} through
\texttt{0xBF} (50 alternatives). Each candidate is authenticated and
encrypted with the same generation key pair. Under the PRP assumption,
each variant produces distinct pseudorandom ciphertext. Termination is
deterministic. The loop is bounded at 51 total attempts (1 primary + 50
alternatives), with an implementation-defined error on exhaustion (the
reference implementation throws \texttt{JackpotException}). For a single
verification key, the idealized estimate for exhausting all 51 attempts
is approximately \(2^{-48 \times 51}\). Multiple enabled keys increase
the per-candidate collision probability.

\subsubsection{Parsing Algorithm}\label{parsing-algorithm-1}

Given a 16-byte UUID \texttt{entityId} and key pair
\((K_{\mathrm{mac}}, K_{\mathrm{aes}})\):

\begin{enumerate}
\def\labelenumi{\arabic{enumi}.}
\tightlist
\item
  Convert the UUID to a 16-byte big-endian byte array (network byte
  order, per RFC 9562).
\item
  Decrypt the 16-byte block using AES-256-ECB with \(K_{\mathrm{aes}}\).
\item
  Check the decrypted plaintext for markers and supported epoch:
  \texttt{guidBytes{[}0{]}\ \textless{}=\ MaxSupportedEpoch},
  \texttt{guidBytes{[}6{]}\ ==\ 0x8D} (exact version match), and
  \texttt{(guidBytes{[}8{]}\ \&\ 0xC0)\ ==\ 0x80} (RFC 9562 §4.1 variant
  range \texttt{0x80}--\texttt{0xBF}). If any check fails, return
  INVALID.
\item
  Verify the decrypted plaintext's MAC using \(K_{\mathrm{mac}}\)
  (extracting bytes 12--15, zeroing the slots, and recomputing the MAC).
  If verification fails, return INVALID.
\item
  Recover the variant byte \(V = \texttt{guidBytes[8]}\). If
  \(V < \texttt{0x8D}\), return INVALID without further key fallback.
\item
  If \(V > \texttt{0x8D}\), perform backward collision-guard
  verification as specified in the Backward Verification Algorithm
  section. If verification fails, return INVALID without further key
  fallback.
\item
  Extract the epoch, entity type, and SKID, and return the result as a
  valid Secure SKEID (\texttt{Secure\ ==\ true}).
\end{enumerate}

\subsubsection{Backward Verification
Algorithm}\label{backward-verification-algorithm}

During parsing, the recovered candidate must first pass its own MAC
verification. When a non-default variant byte \(V\) is recovered from
the decrypted plaintext (\(\texttt{0x8D} < V \leq \texttt{0xBF}\)), the
parse algorithm must verify every preceding variant from \(V-1\) down to
\texttt{0x8D}, inclusive. For each candidate, it reconstructs the SKEID
with the same SKID, epoch, and entity type, recomputes its MAC, encrypts
it with the authenticated key pair, and checks the resulting ciphertext
for a plaintext collision under any enabled verification key. The
reconstruction key pair remains fixed throughout the proof. The parser
rejects the identifier as soon as any preceding candidate does not
collide, without retrying the identifier under another key.

This verifies the complete preceding collision history directly, without
assuming that untrusted input was produced by the generator. For
example, accepting variant \texttt{0x8F} requires collisions at both
\texttt{0x8E} and \texttt{0x8D}. A collision at \texttt{0x8E} alone is
insufficient. Verification requires at most 50 preceding-candidate
checks.

For the default variant (\texttt{0x8D}), parsing performs no predecessor
checks. Under the idealized single-key estimate, generation requires
variant escalation with probability approximately \(2^{-48}\) per
identifier. Epoch filtering can reduce this probability, while
additional enabled verification keys can increase it.

A successfully generated Secure SKEID cannot be accepted as a plaintext
SKEID, provided that generation and parsing use the same verification
keys, epoch-acceptance policy, and plaintext-acceptance checks.

\subsubsection{Auto-Detection Parse
Logic}\label{auto-detection-parse-logic}

Applications operating across both trusted internal and external
boundaries can use auto-detection parsing to accept both plaintext and
encrypted identifiers:

\begin{enumerate}
\def\labelenumi{\arabic{enumi}.}
\tightlist
\item
  Convert the UUID to a 16-byte big-endian byte array (network byte
  order, per RFC 9562).
\item
  If the byte array exhibits a supported epoch
  (\texttt{guidBytes{[}0{]}\ \textless{}=\ MaxSupportedEpoch}) and exact
  primary markers (\texttt{0x8D} at position 6, \texttt{0x8D} at
  position 8), attempt plaintext verification using the Plain SKEID
  Parsing Algorithm. If MAC verification succeeds, return as a plaintext
  SKEID.
\item
  Otherwise, attempt secure parsing using the Secure SKEID Parsing
  Algorithm by decrypting with \(K_{\mathrm{aes}}\) and validating
  markers, MAC, and backward collision proofs.
\item
  If neither path succeeds, return INVALID.
\end{enumerate}

The probability of a random or encrypted byte sequence coincidentally
containing the two marker bytes \texttt{0x8D8D} at the exact positions
is \(2^{-16} = 1/65{,}536\). Requiring one specific 8-bit epoch further
reduces this coincidental match probability to
\(2^{-(16+8)} = 2^{-24} = 1/16{,}777{,}216\). If plaintext MAC
verification fails, parsing falls back to the decryption path. For
generated Secure SKEIDs, the collision guard ensures that plaintext
acceptance fails.

Consequently, endpoints exposed to untrusted networks must avoid
auto-detection or explicitly require the encrypted representation.
Because plaintext SKEIDs expose internal metadata (timestamps, entity
types, topology, and sequence counters) in fixed byte positions,
accepting them at external boundaries permits structure-based
enumeration. An attacker could then craft structured candidates and test
guessed MACs directly against the 32-bit truncation bound, bypassing the
confidentiality and anti-enumeration guarantees of AES-256 encryption.
In the reference implementation, callers enforce this boundary by
selecting \texttt{SourceKnownEntityIdFormat.Secure} rather than
\texttt{Auto}.

\subsubsection{Tier Conversion}\label{tier-conversion}

Implementations support bidirectional conversions between tiers (Figure
1) for already-validated identifiers:

\begin{itemize}
\tightlist
\item
  \textbf{\texttt{ToSecure}:} Converts a plaintext SKEID to a Secure
  SKEID by invoking the Secure SKEID generation algorithm. If the
  identifier is already secure, it is returned unchanged.
\item
  \textbf{\texttt{ToPlain}:} Converts a Secure SKEID to a plaintext
  SKEID. When converting an encrypted identifier whose collision guard
  escalated the variant byte (\texttt{0x8E}--\texttt{0xBF}),
  \texttt{ToPlain} does not expose the raw decrypted block (which would
  fail standard plaintext marker parsing). Instead, it canonicalizes the
  identifier by reconstructing the canonical plaintext SKEID with the
  default variant byte \texttt{0x8D} and a freshly computed BLAKE3 keyed
  MAC under \(K_{\mathrm{mac}}\). If the identifier is already
  plaintext, it is returned unchanged.
\end{itemize}

\subsubsection{Numeric Walkthrough}\label{numeric-walkthrough}

This example reproduces the public test vector from Appendix A of the
draft specification (Kılıç, 2026b). The collision-guard scenario that
follows is hypothetical and is not triggered by this vector.

\textbf{Setup:} Consider a SKID with value \texttt{0x881B3200050C002A}
(signed decimal \texttt{-8639256484514103254}), timestamp = 272,000,000
ticks from epoch (68,000,000 seconds at 250ms precision), app ID = 5,
app instance ID = 3, sequence = 42, entity type \texttt{0x01}, and epoch
\texttt{0x00} beginning on 2025-01-01T00:00:00Z.

The public test key material and derived keys are shown below. Each
hexadecimal string represents 32 raw bytes, not UTF-8 text.
Implementations can use the displayed derived keys directly to reproduce
the wire vector in tests. These public test keys must not be used in
production.

The public master-key value also appears as the 256-bit key-encryption
key in the test vector in RFC 3394 (Schaad \& Housley, 2002, sec. 4.3).
The MAC and AES keys shown below are derived from this master key by the
reference implementation.

\begin{Shaded}
\begin{Highlighting}[]
\NormalTok{Master key:}
\NormalTok{000102030405060708090A0B0C0D0E0F101112131415161718191A1B1C1D1E1F}
\NormalTok{MAC key (K\_mac):}
\NormalTok{5E293E89136745A96B70EB8C8F81CDFCAED177BE5358BC83D3039FB6607FD8FE}
\NormalTok{AES key (K\_aes):}
\NormalTok{4988F97FF724CD086BDFEC83497C3527B3656F35F0911BEEAA6BCE4BB92D3BC7}
\end{Highlighting}
\end{Shaded}

\textbf{Step 1 SKEID Construction:} The 16-byte SKEID buffer is
constructed in big-endian order. Byte 0 receives the epoch
(\texttt{0x00}), bytes 1--4 receive the sign-toggled SKID upper half
(\texttt{0x881B3200} XOR \texttt{0x80000000} = \texttt{0x081B3200}),
byte 5 receives \texttt{0x05}, byte 6 receives the version marker
(\texttt{0x8D}), byte 7 receives the entity type (\texttt{0x01}), byte 8
receives the default variant marker (\texttt{0x8D}), and bytes 9--11
receive \texttt{0x0C002A}. With bytes 12--15 initially zero, the first
four BLAKE3 keyed output bytes are \texttt{492C0E75}. Writing this MAC
into bytes 12--15 produces:

\begin{Shaded}
\begin{Highlighting}[]
\NormalTok{SKEID bytes: 00 08 1B 32 00 05 8D 01 8D 0C 00 2A 49 2C 0E 75}
\NormalTok{SKEID UUID:  00081b32{-}0005{-}8d01{-}8d0c{-}002a492c0e75}
\end{Highlighting}
\end{Shaded}

\textbf{Step 2 Encryption and Collision Check:} Encrypting the 16-byte
plaintext with AES-256-ECB using \(K_{\mathrm{aes}}\) and no padding
produces ciphertext \(C_1\):

\begin{Shaded}
\begin{Highlighting}[]
\NormalTok{Secure bytes: 65 20 68 A4 36 12 CC 4B 8A BB 83 B8 53 DC 67 86}
\NormalTok{Secure UUID:  652068a4{-}3612{-}cc4b{-}8abb{-}83b853dc6786}
\end{Highlighting}
\end{Shaded}

Ciphertext bytes 6 and 8 are \texttt{0xCC} and \texttt{0x8A}, so the
plaintext marker checks fail and the collision guard accepts \(C_1\)
without variant escalation. Parsing either representation recovers SKID
\texttt{-8639256484514103254} and entity type 1. With the same keys,
\texttt{ToSecure} and \texttt{ToPlain} reproduce the two UUIDs above.

\textbf{Hypothetical Collision Scenario:} For a different input or key
pair, suppose the first ciphertext candidate \(C_1\) has a supported
epoch byte, contains \texttt{0x8D} at positions 6 and 8, and has a valid
BLAKE3 keyed MAC at positions 12--15 when interpreted as a plaintext
SKEID. The MAC is computed over a copy of the full 16-byte buffer of
\(C_1\) with bytes 12--15 set to zero. Conditional on the epoch byte
being supported, this marker-plus-MAC coincidence has probability
\(\approx 2^{-48}\) for one verification key.

When the collision guard activates, the plaintext variant byte (position
8) is changed from \texttt{0x8D} to \texttt{0x8E}. The BLAKE3 MAC is
recomputed over the modified plaintext. The new plaintext is encrypted
with the same \(K_{\mathrm{aes}}\), producing ciphertext \(C_2\). Under
the ideal-permutation approximation, the next distinct candidate's
collision probability, including the single accepted epoch, is
approximately \(2^{-56}\). If \(C_2\) passes the check, it becomes the
final Secure SKEID; otherwise, generation continues with the next
variant.

\textbf{Backward Verification:} When a consumer parses \(C_2\) by
decrypting with \(K_{\mathrm{aes}}\), the recovered plaintext reveals
variant byte \texttt{0x8E} (\(> \texttt{0x8D}\)). The parser must verify
that the escalation was legitimate.

\begin{enumerate}
\def\labelenumi{\arabic{enumi}.}
\tightlist
\item
  Reconstruct the SKEID with variant \texttt{0x8D} (replacing
  \texttt{0x8E} and recomputing the MAC).
\item
  Encrypt that reconstruction with \(K_{\mathrm{aes}}\) to obtain
  \(C_1\).
\item
  Check that \(C_1\) has a supported epoch, matching markers, and a
  valid MAC, confirming the collision that justified the escalation.
\item
  For recovered variant \texttt{0x8E}, \texttt{0x8D} is the only
  preceding candidate, so this check covers the complete chain. Higher
  recovered variants require checking every preceding candidate down to
  \texttt{0x8D}.
\end{enumerate}

If the backward check fails (the reconstructed \(C_1\) does not exhibit
the coincidence), the parser rejects the identifier as invalid. This
prevents an attacker from crafting identifiers with arbitrary variant
bytes.

\texttt{IetfDraftTestVectorTests} asserts the master-key derivation,
SKID, MAC, plaintext and ciphertext bytes, UUID strings, and round trips
shown above.

\subsection{Epoch and Extensibility}\label{epoch-and-extensibility}

SKEID byte 0 carries an 8-bit epoch index. Each value selects a
\(2^{31}\)-second window (approximately 68 years), with epoch zero
beginning on 2025-01-01. The 256 epoch values provide approximately
17,421 years of total coverage.

Table 5 illustrates epoch addressing across the full epoch range.

\textbf{Table 5:} Epoch addressing.

{\def\LTcaptype{none} % do not increment counter
\begin{longtable}[]{@{}lll@{}}
\toprule\noalign{}
Epoch Value & Start & End \\
\midrule\noalign{}
\endhead
\bottomrule\noalign{}
\endlastfoot
\texttt{0x00} & January 1, 2025 & January 19, 2093 \\
\texttt{0x01} & January 19, 2093 & February 7, 2161 \\
\(\vdots\) & \(\vdots\) & \(\vdots\) \\
\texttt{0xFF} & 19378 & 19446 \\
\end{longtable}
}

This two-level time addressing gives the identity system a multi-century
lifespan without identifier collisions or sorting degradation. Epoch
boundaries, total coverage, half-epoch boundary year, and sign-bit
sort-order invariants in Table 5 are verified by the
\texttt{PaperEpochAddressabilityTests} unit test suite in the reference
implementation.

\subsection{Defense-in-Depth Security
Architecture}\label{defense-in-depth-security-architecture}

The three-tier identity system employs multiple verification and defence
layers. An attacker attempting to forge a valid identifier must satisfy
all applicable validation checks. AES-256 provides confidentiality and
prevents structure-based enumeration under the PRP assumption. However,
it is not an additional independent forgery-probability factor.

Table 6 enumerates the defense-in-depth security layers and their
contributions to rejecting uniformly random decrypted blocks.

\textbf{Table 6:} Defense-in-depth security layers.

{\def\LTcaptype{none} % do not increment counter
\begin{longtable}[]{@{}
  >{\raggedright\arraybackslash}p{(\linewidth - 4\tabcolsep) * \real{0.2222}}
  >{\raggedright\arraybackslash}p{(\linewidth - 4\tabcolsep) * \real{0.4103}}
  >{\raggedright\arraybackslash}p{(\linewidth - 4\tabcolsep) * \real{0.3675}}@{}}
\toprule\noalign{}
\begin{minipage}[b]{\linewidth}\raggedright
Layer
\end{minipage} & \begin{minipage}[b]{\linewidth}\raggedright
Mechanism
\end{minipage} & \begin{minipage}[b]{\linewidth}\raggedright
Random-block acceptance contribution
\end{minipage} \\
\midrule\noalign{}
\endhead
\bottomrule\noalign{}
\endlastfoot
1. AES-256 Encryption & Confidentiality with PRP over 128 bits &
Prevents structure-based enumeration \\
2. BLAKE3 Keyed MAC & 32-bit truncated & \(2^{-32}\) \\
3. Marker Detection & \texttt{0x8D8D} at fixed positions (default
variant) & \(2^{-16}\) \\
4. Epoch Match & Must match one expected 8-bit epoch or range &
\(2^{-8}\) for single accepted epoch \\
5. Entity Type Match & Must match one expected 8-bit entity type &
\(2^{-8}\) optional \\
6. App ID Match & Must match one expected 7-bit app ID & \(2^{-7}\)
optional \\
7. Existence Probability & Forged SKID must reference a record &
Deployment dependent \\
8. Rate Limiting & Restricts attempts per time period & Implementation
defined \\
\end{longtable}
}

For a uniformly random decrypted block, the default-marker, epoch,
entity-type, app ID, and MAC checks have independent acceptance events.
This allows their acceptance probabilities to be multiplied. For each
key pair, all five validation checks leave \(2^{57}\) plaintext blocks
out of \(2^{128}\). AES permutes these blocks without changing their
number.

For a secure SKEID-only endpoint validating against one key pair in the
key-ring, if validation checks only the 32-bit keyed MAC and the two
fixed marker bytes, the default-variant acceptance probability is
\(2^{-(32+16)} = 2^{-48}\) per uniformly random ciphertext attempt.
Requiring one specific 8-bit epoch reduces it to \(2^{-56}\). Requiring
one specific 8-bit entity type further reduces it to \(2^{-64}\), and
one specific 7-bit app ID reduces it to \(2^{-71}\).

The probabilities above apply specifically to the default-variant path.
Every non-default variant also requires backward collision-guard
verification. Every reconstructed predecessor ciphertext, down to
variant \texttt{0x8D}, must satisfy the plaintext marker, epoch, and MAC
checks. This further restricts acceptance beyond the current variant's
own validation checks.

Record existence checks can reduce acceptance further. Accepting
multiple key pairs, entity types, app IDs, or epochs requires adjusting
the default-variant acceptance probability. These estimates concern
random-ciphertext guessing, not replay or a formal compositional
security proof (see Limitation 4). At endpoints accepting plaintext
SKEIDs, an attacker can set the public fields directly, so these field
checks do not extend the 32-bit MAC forgery bound.

\subsubsection{MAC Truncation Analysis}\label{mac-truncation-analysis}

The following comparison applies the HMAC guidance to BLAKE3 Keyed MAC
and default-variant validation checks solely by analogy. It does not
make the 32-bit BLAKE3 MAC equivalent to a 64-bit MAC or establish
compliance with NIST HMAC guidance.

The SKEID 32-bit MAC truncation matches the minimum MacTag length
specified for HMAC using approved hash algorithms in NIST SP 800-107
(Dang, 2012, sec. 5.3.3). Section 5.3.5 of the same document also
discourages MacTags shorter than 64 bits. However, the 128-bit SKEID
format constrains the space available for the MAC field, as the
remaining bits carry the SKID, entity type, epoch, and marker bytes.
Under the conditions stated in Defense-in-Depth Security Architecture,
the probability that a uniformly random ciphertext passes all Secure
SKEID default-variant validation checks is \(2^{-71}\), lower than the
per-attempt random-tag guessing probability of \(2^{-64}\) for a 64-bit
MacTag.

Section 5.3.5 of the same document provides the general framework for
assessing truncated MAC forgery risk. For a \(\lambda\)-bit MacTag and
an attempt budget of \(q = 2^t\), Equation \eqref{eq:mac-forgery-bound}
gives the idealized tag-guessing bound:

\begin{equation}
\label{eq:mac-forgery-bound}
\Pr[\mathrm{forge}] \leq \min(1, q \cdot 2^{-\lambda})
  = \min(1, 2^{t-\lambda}).
\end{equation}

Applying this estimate by analogy to the 32-bit MAC in plaintext SKEIDs
(\(\lambda = 32\)): if a system permits \(2^{12}\) (4,096) failed
verification attempts before retiring that MAC key from verification,
the forgery likelihood is bounded by \(2^{-20}\), approximately one in a
million, under the idealized tag-guessing model. At 100 rate-limited
attempts per second, this \(2^{12}\) budget is exhausted in roughly 41
seconds.

Secure SKEIDs use AES-256 encryption under the PRP assumption to prevent
structure-based enumeration. At endpoints that reject plaintext SKEIDs,
encryption also prevents direct testing of guessed plaintext MAC values.
Random ciphertext guesses remain subject to validation and rate
limiting.

In December 2022, NIST announced plans to withdraw SP 800-107 Rev.~1
(NIST, 2022). The 2024 Initial Public Draft of NIST SP 800-224 (Turan \&
Brandão, 2024) includes guidance on truncated HMACs. Its recommendations
inform this BLAKE3 analysis by analogy.

\subsection{Key Rotation and Key-Ring
Fallback}\label{key-rotation-and-key-ring-fallback}

A key-ring containing one or more key entries is required. Exactly one
key is designated as the default (active) key at any time. During
routine rotation, applications may retain previous keys for parse
fallback. All new SKEID and Secure SKEID generation uses the current
default key. The reference implementation accepts only the default key
unless the application explicitly enables
\texttt{UseSourceKnownIdKeyFallback} (default \texttt{false}). With
fallback enabled, parsing attempts verification with the current default
key first, then the remaining keys in configured order. Once a candidate
authenticates, its key pair remains fixed for backward verification. A
failed proof rejects the identifier.

Key rotation does not require re-encrypting existing database records
because the 64-bit SKID stored in the database contains no cryptographic
material. Given the SKID and entity type, a new SKEID or Secure SKEID
can be regenerated with the new key at any time. In the event of key
compromise:

\begin{enumerate}
\def\labelenumi{\arabic{enumi}.}
\tightlist
\item
  Set a fresh, uncompromised key as the default.
\item
  Remove the compromised key from verification on every instance.
  Disabling fallback rejects all non-default keys, while removing the
  compromised entry allows other trusted fallback keys to remain
  enabled.
\item
  Reissue affected external identifiers from trusted 64-bit SKIDs and
  entity types as needed. Previously issued identifiers that require the
  removed key no longer authenticate, while the stored 64-bit SKIDs
  remain unchanged.
\end{enumerate}

The application owns key retention and rollout policy. Retaining a
compromised key for compatibility also retains the attacker's ability to
forge accepted identifiers.

\subsection{Reference Implementation}\label{reference-implementation}

The reference implementation is provided as part of the open-source
DRN.Framework (Kılıç, 2026c), organized across four NuGet packages.

The current implementation generates and accepts only encoded epoch
\texttt{0}. Support for nonzero epochs and operational epoch transitions
remains future work.

\begin{itemize}
\tightlist
\item
  \textbf{DRN.Framework.SharedKernel} defines interfaces
  (\texttt{ISourceKnownEntityIdOperations}), base classes
  (\texttt{SourceKnownEntity}), and parsed identity structs
  (\texttt{SourceKnownId}, \texttt{SourceKnownEntityId}).
\item
  \textbf{DRN.Framework.Utils} implements cryptographic operations,
  timestamp management, sequence management
  (\texttt{SequenceManager\textless{}TEntity\textgreater{}} with
  lock-free atomic operations), and bit packing
  (\texttt{SourceKnownIdUtils}, \texttt{SourceKnownEntityIdUtils}).
\item
  \textbf{DRN.Framework.EntityFramework} implements persistence
  integration including \texttt{SourceKnownIdValueGenerator} (EF Core
  value generator for automatic SKID assignment),
  \texttt{DrnSaveChangesInterceptor} (auto-assigns SKID and SKEID on
  insert), \texttt{DrnMaterializationInterceptor} (reconstructs full
  SKEID from stored SKID on database read), and
  \texttt{SourceKnownRepository\textless{}TContext,\ TEntity\textgreater{}}
  (generic repository with SKEID validation and cursor-based
  pagination).
\item
  \textbf{DRN.Framework.Testing} implements integration and unit test
  infrastructure that provides Testcontainers orchestration for
  ephemeral PostgreSQL instances, auto-mocking through data attributes
  (\texttt{{[}DataInline{]}}, \texttt{{[}DataInlineUnit{]}}), and
  convention-based test contexts (\texttt{DrnTestContext},
  \texttt{DrnTestContextUnit}).
\end{itemize}

The specified identifier structure and tier transformations provide a
common contract across domain models, identifier utilities, and
persistence integration. The reference implementation demonstrates this
organization through shared contracts in DRN.Framework.SharedKernel,
generation and parsing in DRN.Framework.Utils, and persistence
integration in DRN.Framework.EntityFramework.

The common contract also enables early validation of expected entity
types and app IDs embedded in the Source Known Identifiers, allowing
mismatched identifiers to be rejected before database access. These
checks support fail-fast behavior and focused tests for incorrect
entity-type or app ID usage.

The implementation uses attribute-based dependency injection
(\texttt{{[}Singleton\textless{}Type\textgreater{}{]}},
\texttt{{[}Scoped\textless{}Type\textgreater{}{]}},
\texttt{{[}Transient\textless{}Type\textgreater{}{]}}) and entity type
attributes (\texttt{{[}EntityType\textless{}TApp\textgreater{}(byte){]}}
for entity type and application partition registration) and integrates
with Domain-Driven Design (Evans, 2003) patterns through the
\texttt{SourceKnownEntity} abstract base class. The testing strategy and
validation evidence for these packages are described in the Validation
and Maturity section.

\textbf{UUID byte-order interoperability:} The SKEID wire layout uses
RFC 9562 network byte order. To avoid the mixed-endian byte-array
representation used by default in .NET \texttt{Guid} APIs, the reference
implementation constructs values with
\texttt{Guid(ReadOnlySpan\textless{}byte\textgreater{},\ bigEndian:\ true)}
and extracts bytes with
\texttt{Guid.TryWriteBytes(guidBytes,\ bigEndian:\ true,\ out\ \_)}.
These conversions preserve the specified byte layout across runtimes.
Secure SKEIDs retain UUID string formatting, but encryption does not
preserve RFC version and variant bits (see Limitation 6).

\subsection{Validation and Maturity}\label{validation-and-maturity}

Active development of the three-tier identity system began in November
2023 as part of the DRN.Framework. The framework has undergone
continuous refinement across versioned releases (v0.1.0 through
v0.10.0). Each release is validated through integration and unit tests
built on DRN.Framework.Testing. Performance is measured separately with
BenchmarkDotNet.

The framework's correctness is validated through multiple test and
analysis tiers.

\begin{itemize}
\tightlist
\item
  \textbf{Unit tests:} In-memory tests covering SKID generation, SKEID
  construction with MAC verification, and Secure SKEID
  encryption/decryption round-trips.
\item
  \textbf{Paper-verification tests:} Dedicated suites
  (\texttt{PaperNumericWalkthroughTests},
  \texttt{PaperEpochAddressabilityTests},
  \texttt{IetfDraftTestVectorTests}) assert the exact numeric values,
  byte layouts, epoch boundaries, lexicographic ordering across epoch
  halves, and sort-order invariants presented in this paper.
  Implementation changes that violate these assertions produce test
  failures.
\item
  \textbf{Integration tests:} Testcontainers-based tests provision
  ephemeral PostgreSQL instances, exercising Entity Framework Core value
  generation, interceptors, and cursor-based pagination with SKID
  ordering.
\item
  \textbf{End-to-end validation:} Test applications validated behavior
  across the three tiers: SKID assignment during entity creation, SKEID
  exposure within trusted application-to-application communication, and
  Secure SKEID generation for external API responses. These applications
  confirm that bidirectional transformations preserve data integrity
  across the complete round-trip.
\item
  \textbf{Static analysis:} GitHub CodeQL and SonarCloud continuously
  scan the codebase for security vulnerabilities, code quality issues,
  and potential bugs. The framework maintains zero critical or
  high-severity findings.
\end{itemize}

The NuGet packages have been published on NuGet.org since the initial
framework release and are available for community review and adoption.
They are published through an automated GitHub Actions CI/CD pipeline
that runs the complete test suite before each release.

\subsection{Reference Benchmark
Environment}\label{reference-benchmark-environment}

The identifier-generation, parsing, and conversion benchmarks under
normal load were conducted using BenchmarkDotNet (Akinshin, 2025)
v0.15.8 with the following hardware and software configuration.

\begin{itemize}
\tightlist
\item
  \textbf{Hardware:} Apple M2, 1 CPU, 8 logical and 8 physical cores
\item
  \textbf{Operating System:} macOS 27.0 (26A428) {[}Darwin 27.0.0{]}
\item
  \textbf{Runtime:} .NET 10.0.12 (10.0.1226.42308), Arm64 RyuJIT
  armv8.0-a
\item
  \textbf{SDK:} .NET SDK 10.0.401
\item
  \textbf{Benchmark configuration under normal load:}
  OutlierMode=RemoveUpper, InvocationCount=262,144, IterationCount=120,
  WarmupCount=120, UnrollFactor=1
\item
  \textbf{Benchmark configuration under saturation:}
  OutlierMode=RemoveUpper, InvocationCount=786,432, IterationCount=40,
  WarmupCount=40, UnrollFactor=1
\end{itemize}

OutlierMode is explicitly set to RemoveUpper, the standard
BenchmarkDotNet throughput default that removes upper statistical
outliers before computing summary statistics.

Mean, Error, and Standard Deviation are reported in nanoseconds (ns) and
rounded to two decimal places throughout the manuscript. Error denotes
the 99.9\% confidence interval (CI) margin calculated by BenchmarkDotNet
over iterations remaining after upper outlier removal. Speedup factors
are calculated by dividing the baseline mean by the operation mean using
the source report values before rounding the result to two decimal
places.

The benchmark configuration under normal load uses 120 measurement
iterations with 262,144 invocations per iteration. This invocation count
matches the per-tick sequence capacity and is chosen to measure
per-operation computational cost within sequence capacity under normal
load. A 260-millisecond \texttt{IterationSetup} delay between iterations
allows the sequence time scope to reset, so each iteration starts from a
clean state.

The saturation benchmark uses \texttt{InvocationCount=786,432} (3 times
the sequence cap), \texttt{IterationCount=40}, and
\texttt{WarmupCount=40} (matching the iteration count), with
\texttt{OutlierMode=RemoveUpper} and \texttt{UnrollFactor=1}. It is
configured to observe backpressure behavior rather than to produce
absolute timings directly comparable with the run under normal load.
These measurements use the same hardware, operating system, runtime, and
SDK listed above.

Both configurations schedule 31,457,280 measurement invocations per
operation (\texttt{InvocationCount\ ×\ IterationCount}), distributing
the same invocation count across different batch sizes to examine
sequence-capacity backpressure.

\subsection{AI-Assisted Research and
Development}\label{ai-assisted-research-and-development}

The following AI tools and large language model (LLM) systems were used
in the research process and development of the reference implementation.

\textbf{Software development, documentation, and testing:} AI tools from
Anthropic (Claude Opus/Sonnet series), Google (Gemini Pro/Flash series),
OpenAI (o1, 4o, GPT-5.6 Sol, GPT-6 Astra), Qwen 3, and DeepSeek R1 were
used across the DRN-Project development lifecycle, including the
Source-Known Identifiers implementation, as agentic coding assistants
for code generation, refactoring, code review, test scaffolding, and
software documentation drafting. CodeRabbit was also used for
AI-assisted code review. Source-Known Identifiers were added to the
DRN-Project roadmap on 19 November 2023. These AI tools were
progressively adopted as they became available over the subsequent
two-plus years of development. All generated code, tests, and
documentation were reviewed, validated, and modified by the author.

\textbf{Literature research:} Google Gemini Deep Research was used to
assist with surveying existing distributed identifier schemes and
locating relevant prior work. All identified sources were independently
verified by the author before citation.

The complete research and reference implementation development history
in the Git version control system is publicly available on GitHub
(https://github.com/duranserkan/DRN-Project), providing an audit trail
of commit progression and automated review interactions.

All AI-assisted outputs were reviewed, validated, and refined by the
author. The architectural design, cryptographic choices, three-tier
trust model, and all technical decisions are solely the work of the
author. The author takes full responsibility for the correctness and
originality of the work.

\section{Results}\label{results}

\subsection{Performance Results}\label{performance-results}

Table 7 presents the BenchmarkDotNet performance measurements for the
measured operations across all three tiers.

\textbf{Table 7:} Generation, parsing, and conversion performance under
normal load.

{\def\LTcaptype{none} % do not increment counter
\begin{longtable}[]{@{}
  >{\raggedright\arraybackslash}p{(\linewidth - 6\tabcolsep) * \real{0.4096}}
  >{\raggedright\arraybackslash}p{(\linewidth - 6\tabcolsep) * \real{0.1446}}
  >{\raggedright\arraybackslash}p{(\linewidth - 6\tabcolsep) * \real{0.1446}}
  >{\raggedright\arraybackslash}p{(\linewidth - 6\tabcolsep) * \real{0.3012}}@{}}
\toprule\noalign{}
\begin{minipage}[b]{\linewidth}\raggedright
Operation
\end{minipage} & \begin{minipage}[b]{\linewidth}\raggedright
Mean (ns)
\end{minipage} & \begin{minipage}[b]{\linewidth}\raggedright
Error (ns)
\end{minipage} & \begin{minipage}[b]{\linewidth}\raggedright
Standard deviation (ns)
\end{minipage} \\
\midrule\noalign{}
\endhead
\bottomrule\noalign{}
\endlastfoot
Random 64-bit integer generation & 177.61 & 2.96 & 9.58 \\
Random UUID V4 generation & 361.94 & 1.94 & 6.31 \\
Random UUID V7 generation & 384.11 & 1.95 & 6.33 \\
SKID generation & 10.85 & 0.64 & 2.09 \\
SKEID generation & 207.28 & 3.10 & 10.08 \\
Secure SKEID generation & 224.62 & 3.90 & 12.67 \\
SKID parsing & 5.35 & 0.41 & 1.34 \\
SKEID parsing & 216.76 & 6.35 & 20.62 \\
Secure SKEID parsing & 250.95 & 3.34 & 10.84 \\
ToPlain & 208.46 & 2.28 & 7.39 \\
ToSecure & 221.23 & 2.98 & 9.62 \\
\end{longtable}
}

\textbf{Note:} Error denotes the 99.9\% confidence-interval margin.
Statistics are calculated over measurement iterations remaining after
upper outlier removal.

Table 8 presents generation performance under saturation. Each batch
contains 786,432 invocations (three times the per-tick sequence
capacity), with a 260-millisecond delay between batches excluded from
measurement.

\textbf{Table 8:} Generation performance under saturation.

{\def\LTcaptype{none} % do not increment counter
\begin{longtable}[]{@{}
  >{\raggedright\arraybackslash}p{(\linewidth - 6\tabcolsep) * \real{0.3766}}
  >{\raggedright\arraybackslash}p{(\linewidth - 6\tabcolsep) * \real{0.1429}}
  >{\raggedright\arraybackslash}p{(\linewidth - 6\tabcolsep) * \real{0.1558}}
  >{\raggedright\arraybackslash}p{(\linewidth - 6\tabcolsep) * \real{0.3247}}@{}}
\toprule\noalign{}
\begin{minipage}[b]{\linewidth}\raggedright
Operation
\end{minipage} & \begin{minipage}[b]{\linewidth}\raggedright
Mean (ns)
\end{minipage} & \begin{minipage}[b]{\linewidth}\raggedright
Error (ns)
\end{minipage} & \begin{minipage}[b]{\linewidth}\raggedright
Standard deviation (ns)
\end{minipage} \\
\midrule\noalign{}
\endhead
\bottomrule\noalign{}
\endlastfoot
SKID generation & 611.22 & 4.82 & 8.57 \\
SKEID generation & 611.23 & 4.99 & 8.88 \\
Secure SKEID generation & 611.67 & 6.33 & 11.25 \\
Random UUID V4 generation & 281.01 & 1.03 & 1.84 \\
Random UUID V7 generation & 302.97 & 1.10 & 1.96 \\
\end{longtable}
}

\textbf{Note:} Error denotes the 99.9\% confidence-interval margin.
Statistics are calculated over measurement iterations remaining after
upper outlier removal.

\subsection{Performance Comparison with
UUID}\label{performance-comparison-with-uuid}

Under normal load, Secure SKEID generation is approximately 1.71 times
faster than UUID V7 (Table 7). Secure SKEID generation incurred
approximately 17.35 ns of additional end-to-end operation cost relative
to plaintext SKEID generation.

SKEID generation is approximately 1.85 times faster than UUID V7 despite
embedding additional metadata (entity type, epoch) and computing a
BLAKE3 keyed MAC within the same 128-bit footprint. Both plaintext and
Secure SKEID generation report zero managed allocation per operation.
Implementations in other managed runtimes may exhibit different
allocation profiles.

SKID generation is approximately 35.39 times faster than UUID V7 and
33.35 times faster than UUID V4. This performance advantage stems from
the deterministic bit-packing approach. SKID generation requires only an
atomic counter increment, a cached timestamp read, and bitwise
operations, with no cryptographic random number generation.

\subsection{Throughput Analysis}\label{throughput-analysis}

The sequence capacity and aggregate sequence capacity bound theoretical
generation throughput. The saturation measurements in Table 8 capture
generation costs including waits at tick boundaries. Because the delay
between batches is excluded, their reciprocals do not represent
continuous sustained throughput.

SKEID and Secure SKEID generation include SKID generation and therefore
incur the same sequence backpressure. The generation speed advantage
over UUID V4 and V7 measured under normal load does not extend to
saturation. Parsing and tier conversion with a provided identifier do
not consume sequence capacity. Their saturation-run measurements are
provided in \texttt{supplementary-saturation-report.html}. All measured
Source-Known operations reported zero managed allocation per operation.
These results support the architectural separation between identifier
generation and identifier processing.

The sequence limits and topology constants underlying the capacity
definitions are checked by the \texttt{PaperThroughputAnalysisTests}
unit test suite in the reference implementation.

\subsection{Storage Comparison}\label{storage-comparison}

Table 1 compares primary key and external identifier sizes across
schemes.

At 8 bytes per SKID versus 16 bytes for UUID, raw key storage
requirements are halved. Compared to KSUIDs at 20 bytes, raw key storage
is reduced by a factor of 2.50 per primary key. Smaller keys can
increase the number of keys per B-tree page and improve cache
utilization. Effects on page splitting depend on the workload and index
configuration. The timestamp-leading layout favors append-heavy
workloads. Cursor-based pagination is natively supported through the
SKID's numeric ordering
(\texttt{WHERE\ id\ \textgreater{}\ :lastId\ ORDER\ BY\ id\ LIMIT\ :pageSize}),
eliminating offset-based pagination overhead. Compared with the
dual-identifier baseline, SKIDs reduce raw identifier column storage
from 24 to 8 bytes per record (66.67\%), deriving the external
representation on demand.

Tier transformations had sub-microsecond mean execution times in the
benchmark environment (Table 7), with no database or network access.
Assuming millisecond-scale latency for I/O-bound identifier resolution,
the measured transformation cost is negligible by comparison.

64-bit SKIDs are timestamp-ordered signed integers within a configured
epoch, and plaintext SKEIDs preserve timestamp order under lexicographic
comparison of their specified big-endian representation. These
properties support an expectation of index locality similar to
conventional time-ordered integer and UUID keys (Davis, Peabody \&
Leach, 2024, sec. 2.1 and 6.11). However, database-engine benchmarks
were not conducted, so improvements in index size, insertion throughput,
and query latency remain unverified.

\section{Discussion}\label{discussion}

\subsection{Interpretation of Results}\label{interpretation-of-results}

The analysis indicates that the six target properties can be distributed
across identity representations under the stated cryptographic
assumptions, provided that deployment coordination preserves uniqueness,
epoch context is retained, and keys and acceptance policies are managed
consistently. In the benchmark environment, generation across all three
tiers was faster on average than UUID V4 and V7 within sequence
capacity, but slower under saturation, with mean generation times
remaining below one microsecond in both conditions.

The measurements distinguish computational cost within sequence capacity
(Table 7) from generation cost under backpressure (Table 8). SKID's
deterministic generation avoids the cryptographic random number
generation required by the measured UUID implementations. Secure SKEID
parsing was the most expensive Source-Known operation under normal load,
while generation costs across the three tiers converged under saturation
because of shared sequence backpressure.

Zero-allocation behavior in core SKID and SKEID operations is important
for managed runtime environments (e.g., .NET, JVM) where garbage
collection pauses can impact tail latency. All measured SKID, SKEID, and
Secure SKEID operations report zero managed allocation per operation,
including secure generation, parsing, and tier conversion.

Expected improvements in database index locality, insertion throughput,
query latency, and garbage-collection-related tail latency remain
untested. Whether a distributed application operates below the aggregate
sequence capacity likewise depends on its workload and available
resources. When a sequence is exhausted, backpressure delays further
generation for the affected application/entity-type pair on that
instance until the next tick, preserving uniqueness within that scope.

\subsection{Implications for Distributed
Applications}\label{implications-for-distributed-applications}

A practical implication is identifier validation at an API gateway or
other trusted edge component provisioned with the required keys. Such a
component can check identifier integrity and expected application and
entity-type metadata without querying the backing database, allowing
invalid identifiers to be rejected before downstream processing. These
checks do not establish record existence or replace authentication and
authorization.

This separation is also useful in eventually consistent systems, where a
receiving service can validate an identifier before the corresponding
record becomes visible in its local read model. Identifier validation
requires no record lookup, while handling replication lag and temporary
record absence remains an application responsibility.

These implications follow from the separation of identifier validation
from record lookup. Their effects on end-to-end latency and throughput
in deployed systems were not evaluated.

\subsection{Comparison with Existing
Work}\label{comparison-with-existing-work}

Compared to UUID V7 (Davis, Peabody \& Leach, 2024), the most directly
comparable recent standard, the three-tier identity system offers origin
metadata, integrity verification (MAC), and confidentiality (encryption)
that UUID V7 lacks. UUID V7 provides millisecond-precision timestamps
while SKID uses 250-millisecond tick precision (see Design Trade-offs
for the rationale and throughput analysis of this choice).

Snowflake (Twitter Engineering, 2012) shares the same compact 64-bit
footprint as SKID, but SKEID adds integrity verification and Secure
SKEID adds confidentiality, neither of which Snowflake provides.
Snowflake's 10-bit machine ID (1,024 machines) is comparable to SKID's
combined 13-bit topology (7-bit app ID + 6-bit app instance ID), but
SKEID adds explicit entity type discrimination. Both require
coordination to assign machine/instance identifiers and both schemes
converge on similar single-epoch durations. However, Snowflake's epoch
is non-extensible while SKEID provides extended epoch coverage. Lastly,
Snowflake uses 41-bit millisecond timestamps while SKID achieves
comparable coverage with a 32-bit 250-millisecond timestamp and an
epoch-half bit.

ULIDs (Feerasta, 2019) require twice the primary key storage (16 bytes
versus 8 for SKID) and carry less embedded metadata. ULIDs use a 48-bit
millisecond timestamp with \textasciitilde8,919 years of coverage from
the Unix epoch, while SKEIDs provide longer coverage through epoch
extension. ULIDs rely on an 80-bit randomness field for probabilistic
uniqueness across independent generators, while SKIDs use deterministic
topology-based partitioning. ULIDs use a 26-character Crockford Base32
encoding that sorts lexicographically, targeting string-first
environments, but lack UUID format compatibility. SKEIDs are valid RFC
9562 Version 8 UUIDs. ULIDs provide neither integrity verification nor
confidentiality.

KSUIDs (Segment, 2023) use 20 bytes per binary identifier (see Table 1).
Source-Known Identifiers provide more compact representations (8/16
bytes) with richer metadata. KSUIDs use second-precision timestamps
while SKIDs use 250-millisecond ticks. KSUIDs provide \textasciitilde136
years coverage with 32-bit seconds and no extension mechanism, whereas
SKEIDs support epoch extension. KSUID's 128-bit random payload trades
storage efficiency for stronger per-identifier collision resistance
without coordination. Source-Known Identifiers achieve uniqueness
through the topology-based partitioning scheme with deployment time
coordination, which guarantees uniqueness by construction rather than by
statistical improbability. Additionally, the three-tier identity system
provides entity type discrimination, integrity verification, and
confidentiality, none of which KSUID offers. Even at the external-facing
tier, Secure SKEIDs (16 bytes) remain more compact than KSUIDs (20
bytes), while the three-tier identity system satisfies all six desired
identifier properties. KSUIDs use a custom 160-bit format with binary
and Base62 representations and are not compatible with standard UUID
parsers.

The original CUID (Elliott, 2023) was deprecated by its maintainer over
security concerns associated with its timestamp-prefixed structure, as
discussed in the Related Work section. Secure SKEIDs address this
metadata exposure directly. The AES-256 encryption layer encrypts the
entire SKEID payload, concealing its timestamp and source metadata from
external consumers without the key.

CUID2 provides probabilistic uniqueness, whereas the three-tier identity
system provides deterministic uniqueness within configured deployment
partitions. For time-based sorting, CUID2 recommends a separate indexed
\texttt{createdAt} column (Elliott, 2026). Source-Known Identifiers can
avoid that additional column when their embedded generation timestamp
satisfies the application's requirements.

Tables 9 and 10 compare the security posture of the three-tier identity
system against standard and alternative identifier schemes.

\textbf{Table 9:} Security comparison of standard identifier schemes.

{\def\LTcaptype{none} % do not increment counter
\begin{longtable}[]{@{}
  >{\raggedright\arraybackslash}p{(\linewidth - 6\tabcolsep) * \real{0.2636}}
  >{\raggedright\arraybackslash}p{(\linewidth - 6\tabcolsep) * \real{0.2000}}
  >{\raggedright\arraybackslash}p{(\linewidth - 6\tabcolsep) * \real{0.2000}}
  >{\raggedright\arraybackslash}p{(\linewidth - 6\tabcolsep) * \real{0.3364}}@{}}
\toprule\noalign{}
\begin{minipage}[b]{\linewidth}\raggedright
Threat
\end{minipage} & \begin{minipage}[b]{\linewidth}\raggedright
SKEID / Secure SKEID
\end{minipage} & \begin{minipage}[b]{\linewidth}\raggedright
UUID V4
\end{minipage} & \begin{minipage}[b]{\linewidth}\raggedright
UUID V7
\end{minipage} \\
\midrule\noalign{}
\endhead
\bottomrule\noalign{}
\endlastfoot
Structure-based enumeration & YES / NO (AES-256) & 122-bit random &
Timestamp + random/monotonic fields \\
Forgery & MAC verification & No detection & No detection \\
Embedded metadata leakage & YES / NO (AES-256) & None (random) &
Timestamp visible \\
Cross-type confusion & Entity type check & Possible & Possible \\
\end{longtable}
}

\textbf{Table 10:} Security comparison of alternative identifier
schemes.

{\def\LTcaptype{none} % do not increment counter
\begin{longtable}[]{@{}
  >{\raggedright\arraybackslash}p{(\linewidth - 10\tabcolsep) * \real{0.2248}}
  >{\raggedright\arraybackslash}p{(\linewidth - 10\tabcolsep) * \real{0.1705}}
  >{\raggedright\arraybackslash}p{(\linewidth - 10\tabcolsep) * \real{0.1550}}
  >{\raggedright\arraybackslash}p{(\linewidth - 10\tabcolsep) * \real{0.1240}}
  >{\raggedright\arraybackslash}p{(\linewidth - 10\tabcolsep) * \real{0.1628}}
  >{\raggedright\arraybackslash}p{(\linewidth - 10\tabcolsep) * \real{0.1628}}@{}}
\toprule\noalign{}
\begin{minipage}[b]{\linewidth}\raggedright
Threat
\end{minipage} & \begin{minipage}[b]{\linewidth}\raggedright
SKEID / Secure SKEID
\end{minipage} & \begin{minipage}[b]{\linewidth}\raggedright
ULID
\end{minipage} & \begin{minipage}[b]{\linewidth}\raggedright
CUID2
\end{minipage} & \begin{minipage}[b]{\linewidth}\raggedright
KSUID
\end{minipage} & \begin{minipage}[b]{\linewidth}\raggedright
Snowflake
\end{minipage} \\
\midrule\noalign{}
\endhead
\bottomrule\noalign{}
\endlastfoot
Structure-based enumeration & YES / NO (AES-256) & TS + 80-bit random &
Hash-based & TS + 128-bit random & Timestamp-ordered \\
Forgery & MAC verification & No detection & No detection & No detection
& No detection \\
Embedded metadata leakage & YES / NO (AES-256) & Timestamp visible &
None (hashed) & Timestamp visible & TS + worker visible \\
Cross-type confusion & Entity type check & Possible & Possible &
Possible & Possible \\
\end{longtable}
}

\textbf{Note:} TS denotes timestamp.

\subsection{Threat Analysis}\label{threat-analysis}

To evaluate the security posture of the three-tier identity system
systematically, we apply a threat analysis informed by the STRIDE
categories (Shostack, 2014) (Spoofing, Tampering, Repudiation,
Information Disclosure, Denial of Service, Elevation of Privilege) to
the three-tier identity model.

\textbf{Spoofing (Identifier Fabrication):} An attacker who intercepts a
Secure SKEID cannot feasibly derive the plaintext SKEID without the
AES-256 key under the PRP assumption. Forging a valid Secure SKEID
requires producing ciphertext that, when decrypted, passes the
applicable marker, MAC, epoch, and expected-identity checks.
Random-ciphertext acceptance probabilities are quantified above under
the conditions stated for Table 6. For plaintext SKEIDs within trusted
environments, resistance to identifier fabrication is limited by the
32-bit MAC and the number of verification attempts.

\textbf{Tampering (Identifier Modification):} Any modification to a
Secure SKEID produces different plaintext upon decryption under the same
AES key, which must still pass the applicable validation checks. For
plaintext SKEIDs, the MAC authenticates the encoded fields, but a
guessed tag for modified fields can still verify with probability
approximately \(2^{-32}\) per attempt under the keyed-MAC assumption.

\textbf{Repudiation:} The three-tier identity system does not provide
non-repudiation in the cryptographic sense because it does not use
digital signatures. However, the embedded topology metadata (app ID, app
instance ID, entity type) and timestamp provide forensic traceability.
For a given SKEID, the encoded application, instance, entity type, and
approximate creation time can be recovered.

\textbf{Information Disclosure:} Plaintext SKEIDs deliberately expose
metadata (timestamp, entity type, topology) within trusted environments
where this information supports routing and validation. For external
consumers, Secure SKEIDs encrypt the entire identifier using
AES-256-ECB, which functions as a pseudorandom permutation on the single
128-bit block. Under the PRP assumption, encryption conceals the encoded
timestamp, entity type, and topology from consumers without the key.
Repeated encryption of the same SKEID plaintext under the same key
produces the same ciphertext, allowing repeated observations of an
identifier to be linked.

\textbf{Denial of Service:} Exhausting the sequence capacity triggers
backpressure for the affected application/entity-type pair on that
instance. Its effect on system-wide availability depends on request
routing and shared resource limits. Parse operations (both encrypted and
plaintext) are bounded-time with no external dependencies, preventing
parse-amplification attacks that require database or network lookups.

\textbf{Elevation of Privilege (Cross-Entity Confusion):} The 8-bit
entity type discriminator enables applications to reject an identifier
for one entity type (e.g., User with entity type 1) submitted as another
(e.g., AdminRole with entity type 5). The unchanged identifier still
passes MAC verification but fails the expected-type comparison.
Applications must separately authorize access to the referenced record,
even when the identifier is valid and matches the expected type (MITRE
Corporation, 2026a).

\subsection{Design Trade-offs}\label{design-trade-offs}

\textbf{32-bit truncated MAC versus full MAC:} The 4-byte MAC is shorter
than typical MACs (16--32 bytes) but is one layer in a defense-in-depth
strategy. The MAC is not the sole security mechanism. It works in
concert with AES encryption, marker detection, entity type matching, and
record existence probability.

\textbf{250-millisecond tick versus millisecond precision:} 64-bit SKIDs
use 250-millisecond tick timestamps (32 bits, four ticks per second)
instead of millisecond-precision (48 bits in UUID V7). This sub-second
quantization provides finer ordering than second-precision schemes while
preserving sufficient bits for application topology and sequence fields
within 64 bits. Applications requiring finer timestamp precision would
need a different allocation of bits or a separate timestamp. Alternative
SKID profiles remain future work.

\textbf{Clock synchronization versus drift protection:} The SKID
generator delegates wall-clock synchronization to the operating system.
Accurate timekeeping through Network Time Protocol (NTP) (Mills et al.,
2010), Precision Time Protocol (PTP) (IEEE, 2020), or equivalent
protocols is an infrastructure responsibility outside the scope of the
identifier generator. The clock drift protection mechanism described in
the SKID specification handles only the consequences of clock jumps or
drifts. External coordination is required for app instance ID assignment
as discussed in Limitation 2. Timestamp monotonicity is guaranteed only
within a running process.

\textbf{Coordination versus probabilistic uniqueness:} The three-tier
identity system requires deployment-time coordination to assign app ID
(7 bits, up to 128 applications) and app instance ID (6 bits, up to 64
instances per application). This stands in contrast to UUID V4, UUID V7,
and CUID2, which achieve uniqueness without any coordination through
cryptographic random number generation. The coordination cost is an
intentional architectural trade-off. Deterministic uniqueness by
construction eliminates the residual collision probability inherent in
probabilistic schemes and allows the compact 64-bit representation that
probabilistic schemes cannot achieve at equivalent uniqueness
guarantees.

\textbf{Topology field sizing:} The 7-bit application field and 6-bit
instance field bound the default profile to 128 applications and 64
instance identifiers per application to keep coordination manageable.
The potential number of pairwise communication relationships
(\(n(n-1)/2\)) motivates this conservative design assumption by analogy
with Brooks's discussion of communication overhead (Brooks, 1995, chap.
2). For 128 applications, this yields 8,128 possible paths.
Well-designed systems mitigate this coordination complexity through
bounded contexts (Evans, 2003) rather than unbounded application
proliferation. Similarly, we assume that beyond approximately 10
concurrently active instances per application, resource contention on
shared infrastructure (databases, message brokers, etc.) becomes the
throughput bottleneck before the identifier generator itself. The 6-bit
field accommodates up to 64 instance identifiers, providing headroom for
operational needs such as rolling deployments and application restarts
triggered by clock drift protection, where the restarting instance must
acquire a new instance identifier to prevent duplicate identifiers.
These are design limits, not empirically established thresholds for
application scale or infrastructure contention. Domain-tailored bit
layouts (see Future Work, item 4) remain available for specialized
deployments requiring different trade-off profiles.

\subsection{Limitations}\label{limitations}

\begin{enumerate}
\def\labelenumi{\arabic{enumi}.}
\item
  \textbf{Uniqueness scope:} 64-bit SKIDs are unique per entity type
  within a deployment and epoch. This requires correct topology
  assignment, clock handling, sequence management, and safe instance
  reuse. Different entity types may share the same SKID value.
\item
  \textbf{Topology limits and coordination requirement:} The default
  profile supports at most 8,192 distinct generators (128 applications ×
  64 instances). Topology assignment (app ID and app instance ID)
  requires external coordination, unlike UUID V4/V7 which achieve
  uniqueness without coordination. The design assumes that a
  configuration or coordination service assigns a different app instance
  ID on every application restart, including ordinary restarts, and
  manages rotation through the 64-value pool. The identifier generator
  does not implement assignment or safe reuse. An app instance ID must
  not be assigned concurrently to multiple processes within the same
  application. Reuse within the same epoch requires the previous holder
  to have stopped and the new generator's timestamp to exceed the
  durably recorded maximum timestamp previously issued under that app
  ID/app instance ID pair. If no identifier can be safely assigned,
  startup must wait or fail.
\item
  \textbf{Key dependency:} SKEIDs and Secure SKEIDs require key material
  for generation and parsing. Loss of key material makes existing SKEIDs
  unparseable, though the underlying 64-bit SKIDs in the database remain
  valid and recoverable.
\item
  \textbf{Security scope:} The security analysis uses STRIDE-based
  threat modeling and quantitative probability analysis within the
  defense-in-depth architecture. It does not provide a formal security
  proof for the composition of AES encryption, the truncated BLAKE3 MAC,
  and the collision guard mechanism. The confidentiality analysis relies
  on modeling AES-256 as a secure pseudorandom permutation. Repeated
  encryption of the same SKEID plaintext under the same key produces the
  same ciphertext, allowing repeated observations of an identifier to be
  linked.
\item
  \textbf{Epoch-scoped storage efficiency:} The 8-byte SKID storage
  efficiency claim applies within a single epoch (\textasciitilde68
  years). The 64-bit SKID encodes the time within the configured epoch
  but does not carry the epoch index itself. A database spanning
  multiple epochs has the following options.

  \begin{enumerate}
  \def\labelenumii{\arabic{enumii}.}
  \item
    A composite key storing the epoch byte separately alongside the
    SKID. This adds per-row overhead that negates the 8-byte claim.
  \item
    Migration to the 16-byte SKEID, which includes the epoch at byte 0.
    This also negates the 8-byte claim.
  \item
    Epoch-partitioned storage, where the table or archive name
    implicitly carries the epoch context (e.g., \texttt{Entity\_Epoch0},
    \texttt{Entity\_Epoch1}). This preserves 8-byte rows by encoding the
    epoch in the partition structure rather than in each record.
  \end{enumerate}
\end{enumerate}

The 8-byte storage representation is suitable when the dataset remains
within one epoch or epoch context is maintained through partitioning.
Long-lived archives and datasets spanning epochs require an explicit
storage and migration strategy. The addressable epoch range alone does
not provide that strategy.

\begin{enumerate}
\def\labelenumi{\arabic{enumi}.}
\setcounter{enumi}{5}
\tightlist
\item
  \textbf{Secure SKEID UUID format compliance:} A Secure SKEID matches
  the UUID length and string format (32 hexadecimal digits in
  \texttt{8-4-4-4-12} grouping) and is accepted by PostgreSQL
  \texttt{uuid} columns and .NET \texttt{Guid} data-type parsers. This
  compatibility is validated by the reference implementation's
  \texttt{UuidVersionAndVariantBatchCompatibilityTests},
  \texttt{SourceKnownEntityIdPostgresUuidCompatibilityTests} and
  \texttt{SourceKnownEntityIdJsonCompatibilityTests} integration tests
  (Kılıç, 2026c). However, AES-256 encryption replaces the plaintext
  Version and Variant marker bytes with pseudorandom ciphertext. The
  encrypted output is not guaranteed to preserve the version and variant
  bits required by RFC 9562 Sections 4.1 and 5.8. Strict RFC 9562
  validators may therefore reject a Secure SKEID. Systems requiring
  strict RFC 9562 compliance at the validator level should accept Secure
  SKEIDs as opaque strings or 128-bit integers.
\end{enumerate}

\subsection{Future Work}\label{future-work}

Future work concerns five areas where the present analysis and
implementation leave open questions.

\begin{enumerate}
\def\labelenumi{\arabic{enumi}.}
\item
  \textbf{Compositional analysis:} A formal analysis is needed to
  establish which security guarantees follow from combining single-block
  AES encryption, the truncated BLAKE3 MAC, and collision-guard
  verification under explicit assumptions. The present threat model and
  probability estimates do not provide such a proof.
\item
  \textbf{AES-CMAC evaluation:} Integrity verification was initially
  planned with HMAC-SHA-256, then shifted toward SHA-3 motivated by
  potential post-quantum security benefits, and subsequently to BLAKE3
  (descended from the SHA-3 finalist BLAKE) after benchmarks confirmed
  its speed advantage on small inputs. However, we discovered that
  single-block hardware-accelerated AES with cached keys may offer
  better performance. We plan to evaluate AES-256-CMAC as specified by
  the CMAC construction in NIST SP 800-38B (Dworkin, 2005). AES-256
  retains a 128-bit block size, matching SKEID's 16-byte authentication
  input. We plan to retain BLAKE3's domain-separated key derivation mode
  (derive\_key) for generating independent encryption and authentication
  keys from master key material in the reference implementation.
\item
  \textbf{Operational lifecycle:} The mechanism for transitioning
  between epochs is not specified. Given that epoch \texttt{0x00} spans
  until approximately 2093 CE, this is left as future work. Future work
  must address how stored identifiers retain their epoch context during
  migration and across long-lived datasets. Key-ring fallback is
  configurable in the reference implementation, but coordinated key
  rollout, revocation, and cross-key ambiguity as the verification key
  set changes remain application responsibilities and subjects for
  further study. Instance identifier assignment and safe reuse across
  restarts likewise require an external coordination mechanism, as
  described in Limitation 2.
\item
  \textbf{Portability and interoperability:} An author-hosted
  \href{https://github.com/duranserkan/DRN-Project/blob/release/v0.10.0/docs/draft-skid-00.md}{draft
  specification (\texttt{draft-kilic-skid-00})} in Internet-Draft format
  (Kılıç, 2026b) is available in the archived DRN-Project v0.10.0
  release, containing algorithms, CDDL definitions, and test vectors.
  Formal IETF submission remains a separate dissemination step.
  Independent implementations in other languages would test the draft's
  interoperability through shared test vectors. Alternative bit
  allocations would require explicit trade-offs among timestamp
  precision, epoch duration, topology capacity, and sequence capacity,
  together with compatibility rules and validation.
\item
  \textbf{Platform-specific UUID compatibility for Secure SKEID:} The
  PostgreSQL and .NET integration tests demonstrate that Secure SKEIDs
  can be stored and parsed without preserving UUIDv8 version and variant
  bits. RFC 9562 recommends treating UUIDs opaquely and avoiding
  unnecessary inspection (Davis, Peabody \& Leach, 2024, sec. 6.12). It
  also defines Nil and Max UUIDs with all 128 bits set to zero or one,
  respectively, and version values 9--15 remain reserved for future
  definitions rather than being permanently excluded from the UUID
  version space (Davis, Peabody \& Leach, 2024, secs. 4.2, 5.9, and
  5.10). These provisions distinguish general UUID handling from
  validation of a particular version, but do not establish RFC
  conformance for arbitrary ciphertext or require parsers to accept it.
  The current results suggest that preserving UUIDv8 bits may also be
  unnecessary on other contemporary platforms. Whether this holds for
  most practical deployments remains to be established. Documenting
  compatibility across platform and library versions is left as future
  work. Where the complete application path accepts Secure SKEIDs,
  techniques such as cycle walking or specialized Feistel constructions
  (Khuong, 2022) may be unnecessary solely for UUID compatibility.
\end{enumerate}

\section{Conclusions}\label{conclusions}

Modern distributed systems face competing demands for storage
efficiency, chronological ordering, and external confidentiality.
Dual-identifier schemas try to address chronological ordering and
external confidentiality at the cost of additional operational and
storage overhead. This paper demonstrates that these goals are not
fundamentally incompatible at the system level when a single entity
identity is represented differently across trust boundaries.

The proposed three-tier architecture aligns SKID (64-bit, database),
SKEID (128-bit, trusted internal), and Secure SKEID (128-bit, external)
with the trust boundaries commonly found in distributed applications.
Deterministic bidirectional transformations enable direct conversion
between tiers. The defense-in-depth security architecture combines
confidentiality, identifier authentication, and application-level
validation, with random-ciphertext acceptance probabilities depending on
the enabled checks.

In the benchmark under normal load within sequence capacity, generation
of SKIDs, SKEIDs, and Secure SKEIDs is faster than generation of either
UUID V4 or UUID V7. Under saturation, backpressure raises Source-Known
generation means above the UUID baselines. All measured identifier
generation, parsing, and conversion operations in both runs have
sub-microsecond mean execution times.

The open-source reference implementation, published as NuGet packages
with integration and unit test coverage, is available for community
review and adoption. A draft specification prepared in Internet-Draft
format is available (Kılıç, 2026b) to support cross-platform
implementation and interoperability testing. Formal submission to the
IETF remains future work.

Within the stated deployment and key-management constraints, the design
illustrates how an entity's identity can remain stable while its
representation changes across trust boundaries to satisfy all six target
properties.

\section*{Data Availability}\label{data-availability}
\addcontentsline{toc}{section}{Data Availability}

The source code for DRN-Project is available on
\href{https://github.com/duranserkan/DRN-Project}{GitHub}. Version
v0.10.0, released on 20 September 2026, is archived on Zenodo (DOI:
\href{https://doi.org/10.5281/zenodo.22863243}{10.5281/zenodo.22863243})
and available under the GitHub tag
\href{https://github.com/duranserkan/DRN-Project/tree/release/v0.10.0}{release/v0.10.0}.
Benchmark data (CSV) and rendered reports (HTML) are included as six
supplementary files covering the benchmarks under normal load,
saturation benchmarks, and hash comparison benchmarks.

\section*{Competing Interests}\label{competing-interests}
\addcontentsline{toc}{section}{Competing Interests}

The author declares no competing interests.

\section*{Funding}\label{funding}
\addcontentsline{toc}{section}{Funding}

This work received no external financial support.

\section*{Author Contributions}\label{author-contributions}
\addcontentsline{toc}{section}{Author Contributions}

Duran Serkan Kılıç conducted the research, conceived the design,
implemented the software, conducted the benchmarks, and wrote the
manuscript.

\section*{Declaration of generative AI and AI-assisted technologies in
the manuscript preparation
process}\label{declaration-of-generative-ai-and-ai-assisted-technologies-in-the-manuscript-preparation-process}
\addcontentsline{toc}{section}{Declaration of generative AI and
AI-assisted technologies in the manuscript preparation process}

During manuscript preparation, the author used Claude Opus 4.6, Gemini
3.1 Pro, Gemini 3.8 Flash, and ChatGPT 6 Astra to assist with drafting,
structuring, and refining sections of this paper, including grammar
correction and editorial refinement. The author reviewed and edited the
AI-assisted content, verified all technical claims against the
implementation, and made all final editorial decisions. The author takes
full responsibility for the content of the manuscript.

\section*{Acknowledgements}\label{acknowledgements}
\addcontentsline{toc}{section}{Acknowledgements}

GitHub CodeQL and SonarCloud provided static analysis and code quality
scanning that strengthened the security posture of the implementation.

\newpage

\section*{References}\label{references}
\addcontentsline{toc}{section}{References}

\protect\phantomsection\label{refs}

\end{document}